\documentclass[12pt]{article}
\pdfoutput=1

\renewcommand{\arraystretch}{1.2}

\usepackage[pdftex]{graphicx}
\graphicspath{{plots/}}

\usepackage{lscape} 
\usepackage{array}

\usepackage{fix-cm}

\usepackage[makeroom]{cancel}
\usepackage{amsmath,amssymb}
\usepackage{slashed}
\usepackage{cite}
\usepackage{pdflscape}
\usepackage{multirow}
\usepackage[thinlines]{easytable}
\usepackage{url}
\usepackage[utf8]{inputenc}
\usepackage{longtable}
\usepackage{booktabs}

\usepackage{braket}
\usepackage{siunitx}
\usepackage{verbatim}
\usepackage[dvipsnames]{xcolor}
\usepackage{xspace}
\usepackage{subfigure}
\usepackage{hyperref}

\newcommand{\SP}[2]{\ensuremath{\mathcal{S}_{#1}^{#2}}}
\newcommand{\FP}[2][P]{\ensuremath{\mathcal{F}^{B\to #1}_{#2}}}
\newcommand{\HP}[2][P]{\ensuremath{\mathcal{H}^{B\to #1}_{#2}}}
\newcommand{\hHP}[2][P]{\ensuremath{\hat{\mathcal{H}}^{B\to #1}_{#2}}}

\newcommand{\tA}[1][ ]{\ensuremath{\tilde{\mathcal{A}}^{#1}}}

\newcommand{\SV}[2]{\ensuremath{\mathcal{S}_{#1}^{#2}}}
\newcommand{\FV}[2][B\to V]{\ensuremath{\mathcal{F}^{#1}_{#2}}}
\newcommand{\HV}[2][B\to V]{\ensuremath{\mathcal{H}^{#1}_{#2}}}

\newcommand{\hHV}[2][B\to V]{\ensuremath{\hat{\mathcal{H}}^{#1}_{#2}}}

\newcommand{\FM}[2][B\to M]{\ensuremath{\mathcal{F}^{#1}_{#2}}}
\newcommand{\HM}[2][B\to M]{\ensuremath{\mathcal{H}^{#1}_{#2}}}
\newcommand{\hHM}[2][B\to M]{\ensuremath{\hat{\mathcal{H}}^{#1}_{#2}}}
\newcommand{\outerF}[2][B\to M]{\ensuremath{\phi^{#1}}_{#2}}

\newcommand{\tV}[2][B\to M]{\ensuremath{\tilde{\mathcal{V}}^{#1}_{#2}}}
\newcommand{\LS}[1][\mu\nu]{\ensuremath{\mathcal{P}^{\tV[]{}}_{#1}}}

\renewcommand{\Im}{\operatorname{Im}}
\renewcommand{\Re}{\operatorname{Re}}

\newcommand{\Disc}{\operatorname{Disc}_{b \bar s}}
\newcommand{\Disccc}{\operatorname{Disc}_{c\bar c}}
\newcommand{\OPE}{\text{OPE}}
\newcommand{\lamkin}{\lambda_{\text{kin}}}

\newcommand{\overbar}[1]{\mkern 1.5mu\overline{\mkern-1.5mu#1\mkern-1.5mu}\mkern 1.5mu}

\newcommand{\para}{\parallel}
\newcommand{\eps}{\varepsilon}
\newcommand{\order}[1]{\mathcal{O}\left(#1\right)}

\newcommand{\GeV}{\ensuremath{\si{\giga\electronvolt}}\xspace}
\newcommand{\MeV}{\ensuremath{\si{\mega\electronvolt}}\xspace}

\newcommand{\LO}{\text{LO}\xspace}
\newcommand{\NLO}{\text{NLO}\xspace}

\newcommand{\refeq}[1]{Eq.~(\ref{eq:#1})}

\newcommand{\refsec}[1]{Section~\ref{sec:#1}}

\allowdisplaybreaks 

\newcommand{\eq}[1]{\begin{equation} #1 \end{equation}}
\newcommand{\eqa}[1]{\begin{eqnarray} #1 \end{eqnarray}}

\newcommand{\Eq}[1]{Eq.~(\ref{#1})}
\newcommand{\Eqs}[2]{Eqs.~(\ref{#1})-(\ref{#2})}
\newcommand{\Reff}[1]{Ref.~\cite{#1}}
\newcommand{\Sec}[1]{Section~\ref{#1}}
\newcommand{\SubSec}[1]{Section~\ref{#1}}
\newcommand{\App}[1]{Appendix~\ref{#1}}
\newcommand{\Tab}[1]{Table~\ref{#1}}
\newcommand{\Fig}[1]{Figure~\ref{#1}}

\newcommand{\K}{\mathcal{K}}
\newcommand{\cO}{\mathcal{O}}

\newcommand{\A}{\mathcal{A}}
\newcommand{\N}{\mathcal{N}}

\renewcommand{\P}{\mathcal{P}}

\newcommand{\av}[1]{\langle #1 \rangle}

\begin{document}

\begin{titlepage}

\vspace*{-2cm}
\begin{flushright}
EOS-2020-01\\
TUM-HEP-1292/20\\
P3H-20-066\\
SI-HEP-2020-27\\[5mm]
\end{flushright}

\vspace{2.2cm}

\begin{center}
\bf
\fontsize{19.6}{24}\selectfont
Non-local matrix elements in $B_{(s)}\to \{K^{(*)},\phi\}\ell^+\ell^-$

\end{center}

\vspace{0cm}

\begin{center}
\renewcommand{\thefootnote}{\fnsymbol{footnote}}
{Nico Gubernari$^{a,b}$, Danny van Dyk$^a$ and Javier Virto$^{a,c}$}
\renewcommand{\thefootnote}{\arabic{footnote}}
\setcounter{footnote}{0}

\vspace*{.8cm}
\centerline{${}^a$\it Technische Universit\"at M\"unchen, Physik Department,}
\centerline{\it James-Franck-Stra\ss{}e 1, 85758 Garching}
\vspace{2.5mm}
\centerline{${}^b$\it Universit\"at Siegen, Naturwissenschaftliche Fakult\"at,}
\centerline{\it Walter-Flex-Stra\ss{}e 3, 57068 Siegen}
\vspace{2.5mm}
\centerline{${}^c$\it Departament de Física Quàntica i Astrofísica, Institut de Ciències del Cosmos,}
\centerline{\it Universitat de Barcelona, Martí Franquès 1, E08028 Barcelona, Catalunya}
\vspace{1.3mm}

\vspace*{.2cm}

\end{center}

\vspace*{10mm}
\begin{abstract}\noindent\normalsize
We revisit the theoretical predictions and the parametrization of non-local matrix elements
in rare $\bar{B}_{(s)}\to \lbrace \bar{K}^{(*)}, \phi\rbrace\ell^+\ell^-$ and
$\bar{B}_{(s)}\to \lbrace \bar{K}^{*}, \phi\rbrace \gamma$ decays.
We improve upon the current state of these matrix elements in two ways.
First, we recalculate the hadronic matrix elements needed at subleading power in the light-cone OPE
using $B$-meson light-cone sum rules. Our analytical results supersede those in the literature.
We discuss the origin of our improvements and provide numerical results for the
processes under consideration.
Second, we derive the first dispersive bound on the non-local matrix elements. It provides
a parametric handle on the truncation error in extrapolations of the matrix elements to large
timelike momentum transfer using the $z$ expansion. We illustrate the power of the dispersive
bound at the hand of a simple phenomenological application.
As a side result of our work, we also provide numerical results for the $B_s \to \phi$
form factors from $B$-meson light-cone sum rules.

\end{abstract}

\end{titlepage}
\newpage 

\renewcommand{\theequation}{\arabic{section}.\arabic{equation}} 

\setcounter{tocdepth}{2}
\tableofcontents


\section{Introduction}
\label{sec:intro}
\setcounter{equation}{0}

Rare $B$ decays mediated by the partonic transition $b\to s \ell^+\ell^-$, such as $\bar B\to \bar K^{(*)}\mu^+\mu^-$ and $\bar B_s\to \phi\mu^+\mu^-$,
are currently the most important probes of the semileptonic operators $[\bar s\Gamma b][\bar\ell \Gamma' \ell]$ in the
Weak Effective Theory (WET)~\cite{Buchalla:1995vs,Aebischer:2017gaw}. These operators may be affected by physics beyond the Standard Model (SM) in a significant way~\cite{Descotes-Genon:2013wba,Beaujean:2013soa,Altmannshofer:2014rta,Descotes-Genon:2015uva,Altmannshofer:2017fio,Aebischer:2019mlg,Alguero:2019ptt,Ciuchini:2020gvn}.
However, robust conclusions about the presence and nature of such New Physics rests upon our ability to calculate the amplitudes with precision and accuracy~\cite{Khodjamirian:2010vf,Khodjamirian:2012rm,
Jager:2012uw,Beaujean:2013soa,Descotes-Genon:2014uoa,Ciuchini:2015qxb,Bobeth:2017vxj,Chobanova:2017ghn}.

To this end, several intrinsically non-perturbative hadronic matrix elements must be calculated reliably.
The main contributions to these amplitudes come from the semileptonic and electromagnetic dipole operators, and they are proportional to hadronic matrix elements of local quark
currents.
These matrix elements, which can be expressed in terms of ``local" form factors, are very similar to the ones appearing in semileptonic decays mediated by charged currents. 
Beyond these terms, there are also contributions from four-quark and chromomagnetic dipole operators, which require the calculation of hadronic matrix elements
of the $T$-product of these local operators and the electromagnetic current~\cite{%
Dimou:2012un,Lyon:2013gba,%
Beneke:2001at,Beneke:2004dp%
}.
These matrix elements can be expressed in terms of ``non-local" form factors
and are considerably more complicated to compute than the local form factors.

The decay amplitudes for any of these processes can be written as\,\footnote{
Here and throughout this work $\bar B$ stands in for either $B^-$, $\bar{B}^0$ or $\bar{B}_s^0$, and $M$ is a hadronic state such that $\bar B\to M$ is a $b\to s$ transition.
The state $M$ may be a single meson or a multiparticle state such as $K^-\pi^+$~\cite{Descotes-Genon:2019bud}.
Our approach works equally for $\bar B \to V \gamma$ decays, as a special case of $\bar B \to V \ell^+\ell^-$.
}
\begin{align}
\label{eq:amplitude}
\A(\bar B\to M \ell^+\ell^-) &=  \frac{G_F\, \alpha\, V^*_{ts} V_{tb}}{\sqrt{2} \pi}\nonumber\\
    &\times \bigg[ (C_9 \,L^\mu_{V} + C_{10} \,L^\mu_{A})\  \FM{\mu} 
        -  \frac{L^\mu_{V}}{q^2} \Big\{  2 i m_b C_7\,\FM{T,\mu}  + 16\pi^2 \HM{\mu} \Big\}   \bigg]\ ,
\end{align}
where $q^2$ is the invariant squared mass of the lepton pair and
$L_{V(A)}^\mu \equiv \bar u_\ell(q_1) \gamma^\mu(\gamma_5) v_\ell(q_2)$ are leptonic currents.
We have included explicitly the effects of the semileptonic operators $\cO_7$, $\cO_9$ and $\cO_{10}$ but suppressed
contributions from other local semileptonic and dipole operators that are not relevant in the SM, as well as from higher-order QED corrections.
Nevertheless, the decomposition in \Eq{eq:amplitude} is exact in QCD. All non-perturbative effects are contained within the local and non-local hadronic matrix elements $\FM{(T),\mu}$ and $\HM{\mu}$, defined as
\eqa{
\label{eq:def-Fmu}
    \FM{\mu}(k, q) &\equiv&  \langle  M(k)|\bar{s}\gamma_\mu P_L\, b|\bar{B}(q+k)\rangle
    \ , \\
    \FM{T,\mu}(k, q) &\equiv& \langle  M(k)|\bar{s}\sigma_{\mu\nu} q^\nu P_R\, b|\bar{B}(q+k)\rangle
    \ , \\
\label{eq:def-Hmu}
    \HM{\mu}(k, q) &\equiv& 
    i\!  \int d^4x\, e^{i q\cdot x}\,
    \av{ M(k) | T\big\{ j_\mu^{\rm em}(x), (C_1\cO_1 + C_2\cO_2)(0) \big\}  | \bar B(q+k)}
    \ ,
}
where $j_\mu^{\rm em} = \sum_{q} Q_q\ \bar q\gamma_\mu q$ with $q=\{u,d,s,c,b\}$. 
In \Eq{eq:def-Hmu} we retained only the terms containing the operators 
\begin{align}
    &
    \cO_1=(\bar s \gamma_\mu P_L T^a c)
    (\bar c \gamma^\mu P_L T^a b)
    &&
    \text{and}
    &&
    \cO_2=(\bar s \gamma_\mu P_L c)
    (\bar c \gamma^\mu P_L b)\,,
    &
\end{align}
which have large Wilson coefficients in the SM.
The contribution of these terms is commonly called the ``charm-loop effect".
The contributions of all the other WET operators are suppressed by small Wilson coefficients and/or by subleading CKM matrix elements.
In the literature, the non-local contributions to \Eq{eq:def-Hmu} are sometimes included through a shift of the Wilson coefficients $C_7$ and $C_9$.
The resulting \emph{effective} Wilson coefficients $C_{7,9}^\text{eff}(q^2)$ become both process- and $q^2$-dependent~\cite{Beneke:2001at}.
In this work we prefer not to use $C_{7,9}^\text{eff}(q^2)$ to keep the non-local contributions explicitly separated from the local ones.

The local and non-local form factors $\FM{\lambda,(T)}(q^2)$ and $\HM{\lambda}(q^2)$ are the invariant functions of a Lorentz decomposition of the matrix elements
\begin{align}
    &
    \FM{(T),\mu}(k, q) \propto \sum_\lambda  \FM{\lambda,(T)}(q^2)\, \SP{\mu}{\lambda}(k, q)\ ,
    &&
    \HM{\mu}(k, q) \propto \sum_\lambda  \HM{\lambda}(q^2)\, \SP{\mu}{\lambda}(k, q)\ .
    &
\end{align}
The structures $\SP{ }{\lambda}$ define our conventions for the form factors and are given in~\App{app:definitions}. 
\\

The non-local contributions $\HM{\mu}$ are currently the main source of theoretical uncertainty in the predictions of $\bar B\to M\ell^+\ell^-$ observables.
All theoretical calculations of  $\HM{\mu}$ rely on some form of Operator Product Expansion (OPE), which allows to expand the non-local $T$-product in terms
of simpler operators. 
Depending on the $q^2$ value, the two relevant OPEs are:
\begin{itemize}
\item \textbf{Local OPE for} $\boldsymbol{|q^2| = \order{m_b^2}}$. 
The OPE is carried out in terms of local operators of the form $\bar{s}(0) D^\alpha \dots D^\omega b(0)$~\cite{Grinstein:2004vb,Beylich:2011aq}.
The matching conditions are known to next-to-leading order in QCD~\cite{Asatryan:2001zw,Greub:2008cy,Ghinculov:2003qd,Bell:2014zya,deBoer:2017way,Asatrian:2019kbk} and the corresponding matrix
elements are related to the local form factors.

\item \textbf{Light-Cone OPE (LCOPE) for}  $\boldsymbol{ 4 m_c^2 - q^2 \gg \Lambda m_b}$.
The OPE is carried out in terms of bilocal light-cone operators of the form $\bar{s}(x) \cO(x,0) b(0)$, with $x^2=0$~\cite{Khodjamirian:2010vf,Khodjamirian:2012rm}.
\end{itemize}

The calculation of the non-local contributions in the region below the open-charm threshold, that is  for $q^2 \lesssim 14 \,\GeV^2$,
then proceeds in three steps:

\begin{enumerate}
    \item Calculation of the LCOPE matching coefficients up to the desired order (both in the QCD coupling and in the LCOPE power counting).
    This calculation must be carried out for $q^2$ values that ensure rapid convergence of the LCOPE, which is the case for $q^2 \lesssim -1 \GeV^2$.

    \item 
    Calculation of the hadronic matrix elements of the operators that emerge in the LCOPE using non-perturbative methods. 

    \item Analytic continuation of the LCOPE results from the point of calculation to $q^2$ values that correspond to semileptonic decays, i.e. for $4 m_\ell^2 \leq q^2 \lesssim 4 M_D^2$.
\end{enumerate}

The matching coefficients of the leading (local) operators of the LCOPE are the same as the ones in the local OPE. 
The next (subleading) order in the LCOPE involves light-cone operators with the insertion of a single soft gluon field. These contributions have been previously considered in Refs.~\cite{Buchalla:1997ky,Khodjamirian:2010vf} where the matching has been computed at leading order in $\alpha_s$.

The matrix elements of the leading operators are related to the local form factors, which have been computed using both lattice QCD and light-cone sum rules (LCSRs) with uncertainties of $10\%$ or less~\cite{Bouchard:2013eph,Horgan:2013hoa,Horgan:2015vla,Gubernari:2018wyi,Straub:2015ica,Descotes-Genon:2019bud}.
The matrix element of the first subleading operator with a soft gluon field has been calculated in the framework of LCSRs with $B$-meson light-cone distribution amplitudes ($B$-LCDAs)~\cite{Khodjamirian:2010vf}.

Finally, the extrapolation to higher values of $q^2$ has been addressed in the literature in a few different ways, ranging from more phenomenological~\cite{Lyon:2014hpa,Ciuchini:2015qxb,Arbey:2018ics,Aaij:2016cbx,Brass:2016efg,Blake:2017fyh}
to more formal~\cite{Khodjamirian:2010vf,Khodjamirian:2012rm,Bobeth:2017vxj}.
The more formal ones involve dispersion relations~\cite{Khodjamirian:2010vf,Khodjamirian:2012rm} or analyticity~\cite{Bobeth:2017vxj}.
Both approaches have the advantage of providing parametrisations that are consistent with QCD, and their parameters can be determined from both theory and data~\cite{Bobeth:2017vxj,Chrzaszcz:2018yza,Mauri:2018vbg}. However, the rate of convergence of these analytic expansions is not well understood, which makes it difficult to assign a truncation error to the approach.

The purpose of this paper is to provide improvements on each of the three steps described above. 
In \SubSec{sec:LCOPE-matching}, we review the calculation of the matching coefficients of the subleading operator in the LCOPE, giving a cleaner representation of the result.
In \SubSec{sec:LCOPE-LCSR}, we recalculate LCSRs for the matrix elements of the non-local subleading operator, employing for the first time a complete set of $B$-LCDAs and updating the values of several crucial inputs.
When putting together these results, we find that the subleading contributions are two orders of magnitude smaller than the previous calculation \cite{Khodjamirian:2010vf}.
Our results significantly reduce the uncertainties on the non-local contributions.
In \Sec{sec:bounds}, we improve on the parametrization of~\Reff{Bobeth:2017vxj} and derive for the first time the dispersive bound on the non-local form factors $\HM{\lambda}(q^2)$. This bound
allows us to constrain the possible effect of truncated terms in the analytic expansion.
Our conclusions follow in \Sec{sec:conclusion}. 
A series of appendices contains details on the definition of the hadronic matrix elements in \App{app:definitions}, the calculation of the matching coefficients
to subleading order in the LCOPE in \App{app:Itensor}, 
the outer functions needed for the dispersive bound in \App{app:fs}, and our results for the $B_s \to \phi$ form factors in \App{app:BsPhiFFs}.


\section{Subleading Contributions in the LCOPE}
\label{sec:subleading-operators}
\setcounter{equation}{0}

We perform a LCOPE to calculate the time-ordered product in \Eq{eq:def-Hmu}.
This is achieved by expanding the charm-quark propagators near the light-cone, i.e. for $x^2\simeq0$.
The first two terms in this expansion are~\cite{Balitsky:1987bk}
\begin{multline}
\label{eq:LCprop}
    \bra{0} T\{c(x)\bar{c}(y)\}\ket{0} = i\int \frac{d^4k}{(2\pi)^4}
    e^{-ik\cdot (x-y)}\frac{\slashed{k}+m_c}{k^2-m_c^2} \\
    -i\!\int\limits_0^1 d u \, G^{\sigma\tau}(ux+(1-u)y)\!\!\int \!\!\frac{d^4k}{(2\pi)^4}
    e^{-ik\cdot (x-y)}
    \frac{(1-u)(\slashed{k}+m_c)\sigma_{\sigma\tau}+u\,\sigma_{\sigma\tau}(\slashed{k}+m_c)}
   {2(k^2-m_c^2)^2}\,,
\end{multline}
where $G_{\sigma\tau}(u x) \equiv g_s t^A G_{\sigma\tau}^{A}(u x)$ is the gluon field.\\\clearpage

The leading power in the LCOPE, which coincides with the leading power of the local OPE, reads\footnote{
    The quantities $\Delta C_7$ and $\Delta C_9$ are the same as in~\Reff{Asatrian:2019kbk}.
}
\begin{multline}
    \label{eq:LOLCOPE}
     i \int d^4x\,e^{i q\cdot x}\,T\big\{ j^\text{em}_\mu(x), (C_1\cO_1 + C_2\cO_2)(0) \big\} 
     \\*
    =\frac{1}{16 \pi^2}  
    \big[
    \left( q_\mu q_\rho- q^2 g_{\mu\rho} \right)
    \Delta C_9\,
    \bar{s}\gamma^\rho P_L b
    + 2 i m_b\, \Delta C_7 \,
    \bar{s}\sigma_{\mu\rho} q^\rho  P_R b
    \big] + \text{higher powers.}
\end{multline}
The matching coefficients $\Delta C_7$ and $\Delta C_9$ have been computed to next-to-leading order in QCD~\cite{Asatryan:2001zw,Greub:2008cy,Ghinculov:2003qd,Bell:2014zya,deBoer:2017way,Asatrian:2019kbk}.
At the leading order, one finds
\begin{align}
    \label{eq:f7LO}
    \Delta C_7(q^2) & = 
    \order{\alpha_s} \,, \\
    \Delta C_9(q^2) & = 
    Q_c
    \left(\frac{4}{3}C_1+C_2\right)
    \Bigg[-
    \frac{8}{9}
    \ln \left(\frac{m_c}{m_b}\right) + \frac{8}{27} +\frac{4}{9}y(q^2)
    \nonumber\\
    \label{eq:f9LO}
    & - \frac{4}{9}\left(2+y(q^2)\right)\sqrt{y(q^2)-1}
    ~\mbox{arctan}\left(\frac1{\sqrt{y(q^2)-1}}\right)
    \Bigg]
    +
    \order{\alpha_s}
    \,,
\end{align}
with $y(q^2)= 4m_c^2/q^2 >1$.

The next-to-leading power term in the LCOPE was discussed for the first time in Ref.~\cite{Khodjamirian:2010vf}. 
In \Sec{sec:LCOPE-matching}, we review the calculation of its matching coefficient at leading order in $\alpha_s$.
In \Sec{sec:LCOPE-LCSR}, we recalculate the corresponding non-local matrix elements using LCSRs with $B$-LCDAs.

\subsection{Matching Condition for the Subleading Operator}
\label{sec:LCOPE-matching}

To obtain the next-to-leading power term in the LCOPE, one has to expand one of the two charm-propagators at next-to-leading power in $x^2$.
This expansion introduces a gluon field $G_{\sigma\tau}(ux)$, as one can see from the second line of \Eq{eq:LCprop}.
Following Ref.~\cite{Khodjamirian:2010vf}, we use the translation operator to rewrite $G_{\sigma\tau}(ux)$ as
\begin{align}
\label{eq:gluon}
    G_{\sigma\tau}(ux)
    =
    e^{-iux\cdot(iD)}G_{\sigma\tau}(0)
    \,.
\end{align}
To make the calculation simpler it is convenient to introduce the light-like vectors $n_\pm^\mu$, which satisfy the following identities:
\begin{align}
    & n_+^2 =  n_-^2 = 0 \,, &
    & n_+ \cdot n_- =2 \,, &
    & n_+^\mu + n_-^\mu = 2 v^\mu \,, &
\end{align}
where $v^\mu$ is the four-velocity of the $B$ meson.
As anticipated in the previous section, to ensure the convergence of the LCOPE one has to impose that $4 m_c^2 - q^2 \gg \Lambda m_b$. 
Hence, we fix
\begin{align}
    & n_- \cdot q \sim m_b \,, &
    & n_+ \cdot q \ll 4 m_c^2 / m_b \,, &
    & q_\perp^\mu \equiv q^\mu - (q \cdot n_+) \frac{n_-^\mu}2 - (q \cdot n_-) \frac{n_+^\mu}2 = 0 \,. &
\end{align}
Now, we can decompose the covariant derivative in \Eq{eq:gluon} in light-cone coordinates:
\begin{align}
    D^\mu = \left(n_+ \cdot D \right) \frac{n_-^\mu}{2}
          + \left(n_- \cdot D \right) \frac{n_+^\mu}{2}
          + D_\perp^\mu
          \,.
\end{align}
As shown in Ref.~\cite{Khodjamirian:2010vf}, the contributions arising from the $(n_- \cdot D)$ are further power-suppressed and hence can be neglected. See also~Refs.\cite{Kozachuk:2018yxf,Melikhov:2019esw}.

We reproduce the form of the subleading operator as in Eq. (3.14) of Ref.~\cite{Khodjamirian:2010vf}:
\begin{align}
    \cO^{\rho\sigma\tau}(\omega_2)
    \equiv
    \bar{s}\gamma^\rho P_L \, \delta\Big(\omega_2 - n_+ \cdot i D\Big) \, G^{\sigma\tau} \, b 
    \,.
\end{align}
This allows us to express the next-to-leading power in the LCOPE as
\begin{multline}
    \label{eq:LCOPE}
     i \int d^4x\, e^{i q\cdot x}\, T\big\{ j_\mu^\text{em}(x), (C_1\cO_1 + C_2\cO_2)(0) \big\}
        =
        \text{local\ contributions} \\*
        + 2 \,Q_c\, C_1^\text{KMPW} 
        \int d\omega_2 \, \tilde{I}_{\mu\rho\sigma\tau}(q, \omega_2 )
        \,
        \cO^{\rho\sigma\tau}(\omega_2)
	    + \, \order{\alpha_s}
	    + \text{higher powers}\,.
\end{multline}
Here $\alpha_s \equiv \alpha_s(\mu_c)$, with $\mu_c$ a perturbative scale, and  $\omega_2$ is the $n_-$ light-cone component of the gluon momentum, which is related to the variable $\omega$
of Ref.~\cite{Khodjamirian:2010vf} through $\omega = \omega_2 / 2$. 
We also abbreviate $C_1^\text{KMPW} = C_2 - \frac{C_1}{2 N_c}$. 
The matching coefficient $\tilde{I}_{\mu\rho\sigma\tau}$ is a four-tensor
that depends on the gluon momentum $\omega_2 n_-$ and the momentum transfer $q$. We reproduce the expression for 
$
I_{\mu\rho\alpha\beta}
    \equiv
    -\frac{1}{2}
    \eps_{\alpha\beta}{}^{\sigma\tau}
    \tilde{I}_{\mu\rho\sigma\tau} 
$
in Eq. (3.15) of Ref.~\cite{Khodjamirian:2010vf} as well. 
Nevertheless, a few comments on  $\tilde{I}_{\mu\rho\sigma\tau}$ are in order:
\begin{itemize}
    \item To leading order in $\alpha_s$ \Eq{eq:LCOPE} only receives contributions from the charm electromagnetic current. 
    Thus, we factored out the charm electric charge in the r.h.s. of \Eq{eq:LCOPE}, rendering the definition of $\tilde{I}_{\mu\rho\sigma\tau}$ consistent with \Reff{Khodjamirian:2010vf}.
    
    \item To leading order in $\alpha_s$, the matching coefficient $\tilde{I}_{\mu\rho\sigma\tau}$ is finite in the limit $\varepsilon \to 0$.
    Hence, it cannot depend explicitly on the renormalization scale $\mu$ to this order. Any residual $\mu$
    dependence emerges only from the use of scale-dependent quantities; here, from the use of the charm quark mass $m_c$.
    This behaviour is reflected in the fact that the matching coefficient of the LCOPE term \emph{at leading power}
    already compensates the running of the semileptonic and electromagnetic dipole coefficients $C_9$ and $C_7$ to order $\alpha_s$.
    
    \item The form in which $I_{\mu\rho\alpha\beta}$ is presented in Ref.~\cite{Khodjamirian:2010vf} is explicitly dependent
    on $\mu$, in apparent contradiction to the previous comment. However, after integrating over $u$, this
    explicit scale dependence is removed. We therefore prefer to present the matching coefficient in such
    a way that the explicit scale dependence does not appear in the first place. Details on the manipulations
    to achieve this are presented in \App{app:Itensor}.
\end{itemize}
Thus, we write the subleading matching coefficient in the form
\begin{align}
    \label{eq:def-Itensor}
    \tilde{I}_{\mu\rho\sigma\tau} (q, \omega_2) 
        & =  \int_0^1 du \int_0^1 dt\,
        \frac{1}{64 \pi^2 ( t(1-t) \tilde{q}^2 - m_c^2)}
        \nonumber\\*
    & \times \bigg(
                4 t(1-t) \left( \tilde{q}_\mu \eps_{\rho \sigma\tau \lbrace \tilde{q}\rbrace} 
                - 2 u \tilde{q}_\tau \eps_{ \mu \rho \sigma \lbrace \tilde{q}\rbrace}
                + 2 u \tilde{q}^2 \eps_{\mu \rho \sigma\tau}
                \right)
                + \tilde{q}^2 \left(1 - 2u\right) \eps_{\mu \rho \sigma\tau}
            \bigg)\,.
\end{align}
Here, we approximate the square of the momentum $\tilde{q}^\mu = q^\mu - v^\mu u \omega_2/2 $ as $\tilde{q}^2 \simeq q^2 - u \omega_2 m_b$ and adopt
the convention $\eps_{0123} = +1$. 
Even though \Eq{eq:def-Itensor} has a slightly different form with respect to the matching coefficient presented in \Reff{Khodjamirian:2010vf},
we emphasize that the two results are analytically equivalent. 
The form \Eq{eq:def-Itensor} has also been used to calculate the numerical results of \Reff{Khodjamirian:2010vf} in a non-public \texttt{Mathematica} code\,\footnote{%
    We are very grateful to Yu-Ming Wang for sharing this code with us.
}. 
Therefore, we fully confirm the analytic results of~\Reff{Khodjamirian:2010vf} for the tensor-valued matching coefficient due to soft-gluon interaction at subleading power in the LCOPE.

\subsection{Calculation of the Non-Local Hadronic Matrix Elements}
\label{sec:LCOPE-LCSR}

We proceed to calculate the hadronic matrix elements of the operator 
\begin{align}
    \tilde{\cO}_\mu (q) \equiv \int d\omega_2 \, \tilde{I}_{\mu\rho\sigma\tau}(q, \omega_2)\, \cO^{\rho\sigma\tau}(\omega_2)
\end{align} 
using LCSRs with $B$-LCDAs.
To derive the sum rule, we start defining the  correlator
\begin{align}
    \label{eq:Pi-LCSR}
	\Pi^\Gamma_{\mu\nu}(k, q)
	    = i \, \int d^4y \, e^{i k\cdot y} \bra{0} T \left\lbrace j_\nu^{\Gamma}(y), \tilde{\cO}_\mu(q) \right\rbrace \ket{\bar{B}(q+k)}\,,
\end{align}
where $j_\nu^{\Gamma} \equiv \bar{q}_1\, \Gamma_\nu s$ is the interpolating current, with $\bar{q}_1=\{\bar u,\bar d,\bar s\}$ depending on the decay channel.
The Dirac structure $\Gamma_\nu$ of the interpolating current is chosen such that the respective $M$-to-vacuum matrix element does not vanish. We use
\begin{equation}
    \Gamma_\nu= \begin{cases}
        \gamma_\nu\gamma_5  & \text{for $M = K$}\\
        \gamma_\nu          & \text{for $M = K^*,\phi$}
    \end{cases}\,.
\end{equation}

The light-cone sum rule calculation of the matrix elements of $\tilde{\cO}_\mu(q)$ is based on a LCOPE
of the correlator \Eq{eq:Pi-LCSR} in the framework of heavy quark effective theory. 
Here, the momentum $q$ is fixed according to the considerations in the previous section.
For a fixed $q$, we now need to ensure that $k^2$ is chosen appropriately, i.e. that the expansion of \Eq{eq:Pi-LCSR} is
dominated by bi-local operators with light-cone dominance. We find that for $-k^2 \sim \text{a few}\,\Lambda^2$
the correlator is dominated by contributions at light-like distances 
$ y^\mu\simeq (y\cdot n_-) \frac{n_+^\mu}{2}$ such that $y\cdot q \ll (y \cdot n_-) 4m_c^2/m_b$ \cite{Khodjamirian:2010vf}. 
Therefore, we can expand the strange quark propagator near the light-cone as well, keeping only the leading power term.
We obtain
\begin{multline}
    \label{eq:Pi-LCSR-OPE}
	\Pi^\Gamma_{\mu\nu}(k, q)
	    = \int d\omega_2 \, \int d^4y \, \int d^4 p' \, e^{i (k - p') \cdot y}
	        \tilde{I}_{\mu\rho\sigma\tau}
	        \left[
	            \Gamma_\nu \, \frac{\slashed{p}' + m_s}{m_s^2 - p'^2} \, \gamma^\rho P_L
	        \right]_{ab}
	        \\*
	 \times\bra{0} \bar{q}_1^a(y) \delta\left[\omega_2 - i n_+ \cdot D\right] G^{\sigma\tau} h^b_v(0) \ket{\bar{B}(v)}\,,
\end{multline}
where $a$, $b$ are spinor indices. 

The non-local $B$-to-vacuum matrix elements can be expressed in terms of $B$-LCDAs.
While in the sum rules of local form factors the leading contribution involves only two-particle $B$-LCDAs, the leading contribution in \Eq{eq:Pi-LCSR-OPE} comes from the three-particle $B$-LCDAs. 
The three-particle distribution amplitudes of the $B$-meson can be defined as~\cite{Geyer:2005fb,Braun:2017liq}
\begin{align}
    \bra{0} \bar{q}_1^a(y) \delta & \left[ \omega_2 - i n_+ \cdot D \right] G_{\sigma\tau} h_v^b(0) \ket{\bar{B}(v)}\Big|_{y\simeq n_+, y^2\simeq 0}
    \nonumber\\*
    & =
    \frac{f_B M_B}{4} \int_0^\infty d\omega_1 e^{-i \omega_1 v\cdot y}\,    
    \bigg\{(1+\slashed{v})\bigg[
        (v_\sigma\gamma_\tau-v_\tau\gamma_\sigma)[\psi_A-\psi_V] 
        -i\sigma_{\sigma\tau}\psi_V  
    \nonumber\\*
    & + (\partial_\sigma v_\tau-\partial_\tau v_\sigma) \overbar{X}_A
      - (\partial_\sigma\gamma_\tau-\partial_\tau\gamma_\sigma) [\overbar{W} + \overbar{Y}_A]
      + i\epsilon_{\sigma\tau\alpha\beta}\partial^\alpha v^\beta\gamma_5 \overbar{\tilde{X}}_A 
    \nonumber\\*
    & - i\epsilon_{\sigma\tau\alpha\beta}\partial^\alpha \gamma^\beta\gamma_5 \overbar{\tilde{Y}}_A
      - (\partial_\sigma v_\tau-\partial_\tau v_\sigma)\slashed{\partial} \overbar{\overbar{W}}
      + (\partial_\sigma \gamma_\tau-\partial_\tau \gamma_\sigma)\slashed{\partial} \overbar{\overbar{Z}}
    \bigg]\gamma_5\bigg\}^{ba}
    \,,
    \label{eq:3ptLCDAs}
\end{align}
where we have suppressed the arguments of the $B$-LCDAs, e.g. $\psi_A \equiv \psi_A(\omega_1, \omega_2)$, for brevity.
The derivatives, which act only on the hard-scattering kernel, are abbreviated as $\partial_\mu \equiv \partial / \partial l^\mu$, with $l^\mu=(\omega_1 + u \omega_2)v^\mu$.
In addition, we introduced the shorthand notation
\begin{equation}
\begin{aligned}
    \overbar{\psi}_{\text{3p}}(\omega_1, \omega_2)
        & \equiv \int_0^{\omega_1} d\eta_1 \, \psi_{\text{3p}}(\eta_1, \omega_2)\,,\\
    \overbar{\overbar{\psi}}_{\text{3p}}(\omega_1, \omega_2)
        & \equiv \int_0^{\omega_1} d\eta_1 \, \int_0^{\eta_1} d\eta_2 \, \psi_{\text{3p}}(\eta_2, \omega_2)\,,
\end{aligned}
\end{equation}
where $\psi_\text{3p}$ represents any of the three-particle $B$-LCDAs appearing in \Eq{eq:3ptLCDAs}.
In Ref.~\cite{Khodjamirian:2010vf} only the $B$-LCDAs $\psi_A$, $\psi_V$, $X_A$, and $Y_A$ have been taken into account.

The distribution amplitudes of \Eq{eq:3ptLCDAs} have no definite twist, which is defined as the difference between the dimension and the spin of the corresponding operator.
It is important to express the $B$-LCDAs in \Eq{eq:3ptLCDAs} in terms of $B$-LCDAs with definite twist, to ensure a consistent power counting.
The relations between the twist basis and the basis of \Eq{eq:3ptLCDAs} are given in Ref.~\cite{Braun:2017liq}.
Inverting these relations, one obtains (using the same nomenclature as in Ref.~\cite{Gubernari:2018wyi})
\begin{align}
    \psi_A(\omega_1, \omega_2) & = 
    \frac{1}{2}\, [ \phi_3 + \phi_4                                             ](\omega_1, \omega_2) 
    \,, \nonumber\\
    \psi_V(\omega_1, \omega_2) & = 
    \frac{1}{2}\, [-\phi_3 + \phi_4                                             ](\omega_1, \omega_2) 
    \,, \nonumber\\
    X_A(\omega_1, \omega_2) & = 
    \frac{1}{2}\, [-\phi_3 - \phi_4 + 2\psi_4                                   ](\omega_1, \omega_2) 
    \,, \nonumber\\
    Y_A(\omega_1, \omega_2) & = 
    \frac{1}{2}\, [-\phi_3 - \phi_4 +  \psi_4 - \psi_5                          ](\omega_1, \omega_2) 
    \,, \nonumber\\
    \tilde{X}_A(\omega_1, \omega_2) & = 
    \frac{1}{2}\, [-\phi_3 + \phi_4 - 2\chi_4                                   ](\omega_1, \omega_2) 
    \,, \nonumber\\
    \tilde{Y}_A(\omega_1, \omega_2) & = 
    \frac{1}{2}\, [-\phi_3 + \phi_4 -  \chi_4 + \chi_5                          ](\omega_1, \omega_2) 
    \,, \nonumber\\
    W(\omega_1, \omega_2)           & = 
    \frac{1}{2}\, [ \phi_4 - \psi_4 -  \chi_4 + \tilde{\phi}_5 + \psi_5 + \chi_5](\omega_1, \omega_2) \,, \nonumber\\
    Z(\omega_1, \omega_2)           & = 
    \frac{1}{4}\, [-\phi_3 + \phi_4 - 2\chi_4 + \tilde{\phi}_5 + 2 \chi_5 - \phi_6](\omega_1, \omega_2) \,,\nonumber
\end{align} 
where the subscripts $3,\,4,\,5,\,6$ indicate the twist of the respective $B$-LCDA.
We include in our calculation contributions up to twist four with the models given in Section 5.1 of Ref.~\cite{Braun:2017liq}.
The models for $B$-LCDAs of twist five or higher are not currently known.
However, they ``are not expected to contribute to the leading power corrections $\order{1/M_B}$ in $B$ decays''~\cite{Braun:2017liq}.
\\

To proceed with the extraction of our sum rule, we need to obtain the hadronic dispersion relation of the correlator (\ref{eq:Pi-LCSR}).
This can be done by inserting a complete set of states with the appropriate flavour quantum numbers in between the interpolating current $j_\nu^{\Gamma}$ and the operator $\tilde{\cO}_\mu$: 
\begin{align}
    \label{eq:Pi-LCSR-had}
	\Pi^\Gamma_{\mu\nu}(k, q)
	    & =\sum_\lambda \frac{
	        \bra{0} j_\nu^{\Gamma} \ket{M(k,\eta(\lambda))} \bra{M(k,\eta(\lambda))} \tilde{\cO}_\mu(q) \ket{\bar{B}(q + k)} }{M_M^2-k^2} 
	      + \int_{s_h}^\infty d s\frac{\rho_{\mu\nu}^\Gamma(s,q^2)}{s-k^2}
	\,,
\end{align}
where the sum runs over all the possible polarizations.
The function $\rho_{\mu\nu}^\Gamma(s,q^2)$ is the spectral density, which encodes the information about the continuum as well as the exited states, while $s_h$ denotes the continuum states threshold. 
The matrix elements of the interpolating current are expressed in terms of the decay constants $f_P$ for $P=K$
and $f_V$ for $V=K^*,\phi$:
\begin{equation}
\begin{aligned}
    \label{eq:decconst}
	\bra{0} \bar{q}_1 \gamma_\nu\gamma_5 s \ket{P(k)}       & = i f_P k_\nu\,,\\
	\bra{0} \bar{q}_1 \gamma_\nu         s \ket{V(k, \eta)} & = i f_V \eta_\nu M_V\,.
\end{aligned}
\end{equation}
The matrix elements of the operator are decomposed in terms of the scalar valued functions $\tV{\lambda}$:
\begin{align}
    \label{eq:non-locFFBK}
	\bra{P(k)} \tilde{\cO}_\mu(q) \ket{\bar{B}(q + k)}
	& =  M_B^2\, \SP{\mu}{0}\, \tV[B \to P]{0}
	\,,\\*
	\bra{V(k, \eta)} \tilde{\cO}_\mu(q) \ket{\bar{B}(q + k)}
	& = 
	M_B^2\, \eta^{*\alpha}\left[
        \SP{\alpha\mu}{\perp} \tV[B \to V]{\perp} - \SP{\alpha\mu}{\para} \tV[B \to V]{\para} - 
        \SP{\alpha\mu}{0} \tV[B \to V]{0}\right]
	\,.
    \label{eq:non-locFFBKstar}
\end{align}
where the definitions of the Lorentz structures $\SP{}{\lambda}$ are given in \App{app:definitions}.
The hadronic dispersion relation for the correlator (\ref{eq:Pi-LCSR}) is  finally obtained by plugging \Eqs{eq:decconst}{eq:non-locFFBKstar} into \Eq{eq:Pi-LCSR-had}.

It is worth emphasizing that the functions $\tV[B \to M]{\lambda}$ are the next-to-leading power contributions in the LCOPE to the non-local form factors $\HV[B \to M]{\lambda}$ that are defined as
\begin{align}
    \HP{\mu}
        & = M_B^2\, \SP{\mu}{0}\, \HP{0} \,,
    \\
    \HV{\mu}
        & = M_B^2\, \eta^{*\alpha}\left[
        \SP{\alpha\mu}{\perp} \HV{\perp} - \SP{\alpha\mu}{\para} \HV{\para} - 
        \SP{\alpha\mu}{0} \HV{0}\right]\,.
\end{align}
The expressions for the non-local form factors $\HV[B \to M]{\lambda}$ then read
\begin{align}
    \label{eq:masterNLFF}
    \HM{\lambda}&=
    -\frac{1}{16\pi^2}
    \left(
        \frac{q^2}{2M_B^2}
        \Delta C_9\, \FM{\lambda}
        + \frac{m_b}{M_B} \Delta C_7\,
        \FM{\lambda,T}
    \right)
    +
    2 \,Q_c\, C_1^\text{KMPW}
    \tV{\lambda}
    + 
    \dots
    \,,
\end{align}
where the ellipses stands for higher powers in the LCOPE and spectator scattering interactions.
The definitions of the form factors $\FM{\lambda}$ are given in \App{app:definitions} as well.
\\

The last step in the calculation of the sum rule is to match the LCOPE result of \Eq{eq:Pi-LCSR-OPE} into the hadronic representation.
To get rid of the contributions of the excited and continuum states of \Eq{eq:Pi-LCSR-had}, we exploit the semi-global quark-hadron duality approximation. 
We also perform a Borel transform to reduce the impact of potential quark-hadron duality violations.
In order to isolate the individual contributions of the functions $\tV{\lambda}$, we select a suitable Lorentz structure \LS.
Hence, we decompose the two-point functions in terms of scalar-valued functions $\Pi^{\tV[]{}}(k^2, q^2)$:
\begin{align}
    \Pi^\Gamma_{\mu\nu}(k, q) \equiv \sum_{\tV[]{}} \LS(k, q) \, \Pi^{\tV[]{}}(k^2, q^2)\,.
\end{align}
The LCOPE results for the functions $\Pi^{\tV[]{}}$ can always be written in the form
\begin{align}
    \Pi^{\tV[]{}}(k^2, q^2)
        = f_B M_B \sum_n \int_0^\infty d\sigma \frac{J^{\tV[]{}}_n(s, q^2)}{\left[k^2 - s(\sigma,q^2)\right]^n}\,,
\end{align}
where we have introduced the new variable $\sigma = \omega_1 / M_B$ and  defined
\begin{align}
    s(\sigma,q^2)=\sigma M^2_B +\frac{m_s^2-\sigma q^2}{1-\sigma}\,.
\end{align}

For ease of comparison, we give our results in the basis $\tA[],\,\tV[]{1},\,\tV[]{2},\,\tV[]{3}$ as in~\Reff{Khodjamirian:2010vf} instead of $\tV[B\to P]{0},\,\tV[B\to V]{\perp},\,\tV[B\to V]{\para},\,\tV[B\to V]{0}$.
The relations between these two bases read
\begin{equation}
    \begin{aligned}
    \tV[B\to P]{0} & = - \frac{q^2}{2 M_B^2} \tA[]
    \,,\\
    \tV[B\to V]{\perp}  & = \frac{\sqrt{\lamkin}}{\sqrt{2} M_B^3} \, \tV[]{1}
    \,,\\
    \tV[B\to V]{\para} & = -\sqrt{2}\,\frac{M_B^2 - M_V^2}{M_B^3} \,  \tV[]{2}
    \,,\\
    \tV[B\to V]{0}     & =
    -\frac{q^2}{2 M_B^4 M_V}\, 
    \left[(M_B^2 + 3 M_V^2 - q^2)\tV[]{2} - \frac{\lamkin}{M_B^2 - M_V^2} \tV[]{3} \right]\,,
    \end{aligned}
\end{equation}
where $\lamkin \equiv \lambda(M_B^2,M_V^2,q^2)$ is the K\"all\'en function.
For convenience, we have also introduced the function
\begin{align}
    \tV[]{23} = 
        \tV[]{2} 
        + \frac{M_V^2 + q^2 - M_B^2}{M_B^2 - M_V^2} \,\tV[]{3} 
    \,.
\end{align}

\begin{table}[t]
    \centering
    \setlength{\tabcolsep}{12pt}
    \begin{tabular}{l c c c@{}}
        \toprule
          $\tV[]{}$             & $\LS$              & $N^{\tV[]{}}$  & $K_2^{\tV[]{}}$\\
        \toprule
        $\tA[]$     &  $q_\mu k_\nu$     & $-\frac{i}{2}$ & $f_P (M_B^2 - M_P^2 - q^2)$\\
        \midrule
        $\tV[]{1}$  & $\eps_{\mu\nu kq}$ & $+1$           & $f_V M_V$\\
        $\tV[]{2}$  & $g_{\mu\nu}$       & $-i$           & $f_V M_V (M_V^2 - M_B^2)$\\
        $\tV[]{23}$ & $q_\mu q_\nu$      & $-i$           & $f_V M_V$\\
        \bottomrule
    \end{tabular}
    \caption{ \it
    Overview of the quantities $\tV[]{}$ extracted from the correlation functions, their corresponding Lorentz structures $\LS$, and their normalization
    factors $N^{\tV[]{}}$ and $K_2^{\tV[]{}}$. See the text for details.}
    \label{tab:LCSR-overview}
\end{table}

We can now write down the sum rule for any of the quantities $\tV[]{}=\tA[],\,\tV[]{1},\,\tV[]{2},\,\tV[]{23}$, which reads
\begin{align}
    \tV[]{}
        & = -\frac{f_B M_B}{K^{\tV[]{}}} \sum_{n=1}^{\infty}\Bigg\{(-1)^{n}\int_{0}^{\sigma_0} d \sigma \;e^{(-s(\sigma,q^2)+M^2_{P,V})/M^2} \frac{1}{(n-1)!(M^2)^{n-1}}I_n^{\tV[]{}}\nonumber\\*
        & - \Bigg[\frac{(-1)^{n-1}}{(n-1)!}e^{(-s(\sigma,q^2)+M^2_{P,V})/M^2}\sum_{j=1}^{n-1}\frac{1}{(M^2)^{n-j-1}}\frac{1}{s'}
        \left(\frac{d}{d\sigma}\frac{1}{s'}\right)^{j-1}I_n^{\tV[]{}}\Bigg]_{\sigma=\sigma_0}\Bigg\rbrace\,,
        \label{eq:masterformula}
\end{align}
with $I_n^{\tV[]{}} \equiv J_n^{\tV[]{}} / N^{\tV[]{}}$.
We abbreviate $\sigma_0\equiv \sigma(s_0,q^2)$, 
$s'(\sigma,q^2) \equiv d s(\sigma,q^2)/d \sigma$, and the differential operator
\begin{equation}
    \left(\frac{d}{d\sigma}\frac{1}{s'}\right)^{n} I(\sigma) \equiv
    \left(\frac{d}{d\sigma}\frac{1}{s'}\left(\frac{d}{d\sigma}\frac{1}{s'}\dots I(\sigma)\right)\right)\,.
\end{equation}
Here $s_0$ is the effective threshold $s_0$ of the sum rule, which differs in general from the continuum threshold $s_h$.
The functions $I_n^{\tV[]{}}$ can be represented as integrals of the three-particle $B$-LCDAs
\begin{align}
    I_n^{\tV[]{}}(\sigma,q^2)
    & = \frac{1}{(1 - \sigma)^n}
        \int \displaylimits_{0}^{\infty}d\omega_2
        \int \displaylimits_{0}^{1}du
        \int \displaylimits_{0}^{1}dt
        \sum_{\psi_\text{3p}}  
        \sum_{r=0}^2  
        \left(\frac{\omega_2}{M_B} \right)^r
        C^{(\tV[]{},\psi_\text{3p})}_{n,r}(\sigma,u,t,q^2)\, 
        \psi_\text{3p} (\sigma M_B, \omega_2) 
        \, ,
\label{eq:CoeffFuncs3pt}
\end{align}
for $\psi_\text{3p}=\phi_3,\phi_4,\psi_4,\chi_4$.
The factors $K^{\tV[]{}}$ in  \refeq{masterformula} consists of a universal part and a structure dependent part:
\begin{align}
    K^{\tV[]{}}(q^2, t,  u, \omega_2) = K_1(q^2,t, u, \omega_2) K_2^{\tV[]{}}(q^2)
    \,,
\end{align}
where
\begin{equation}
    K_1(q^2, t, u, \omega_2) = 8\pi^2 \left[m_c^2-t(1-t)\left(q^2 - u \omega_2 M_B\right)\right]\,.
\end{equation}
The quantities $\LS$, $N^{\tV[]{}}$, and $K_2^{\tV[]{}}$ are listed in \Tab{tab:LCSR-overview}, while the coefficients $ C^{(\tV[]{},\psi_\text{3p})}_{n,r}$ of \Eq{eq:CoeffFuncs3pt} are provided in an ancillary \texttt{Mathematica} file. 
\\

We do not expect to find full agreement between our results  and those of
Ref.~\cite{Khodjamirian:2010vf} for the matching coefficients $C_{n,r}^{\tV[]{},\psi_{3p}}$. The reason is that our results are expressed in terms of the full set of three-particle $B$-LCDAs as discussed
in Ref.~\cite{Braun:2017liq}, while the results of Ref.~\cite{Khodjamirian:2010vf} use an incomplete set of Lorentz structures
and $B$-LCDAs.
For the calculation of local form factors, this issue is not numerically relevant, since the three-particle contributions
are numerically small compared to the leading twist and even the next-to-leading twist two-particle contributions;
see also the discussion in Ref.~\cite{Gubernari:2018wyi}.
For this particular LCSR calculation,
the two-particle contributions are absent and hence the three-particle contributions are numerically leading.

Our main results can be summarized as follows:
\begin{itemize}
    \item Restricting our results to the same set of Lorentz structures and independent three-particle $B$-LCDAs as in \Reff{Khodjamirian:2010vf},
    we find full agreement with the results of that paper.
    
    \item Using the full set of three-particle $B$-LCDAs, the thresholds setting procedure of~\Reff{Imsong:2014oqa} produces results that
    are compatible with the thresholds obtained for the local form factors in \Reff{Gubernari:2018wyi}.
    This is not the case when restricting our analytical expression to the subset of Lorentz structures as discussed in the previous point.
    
    \item 
    Our final results are \emph{one order of magnitude smaller} than in \Reff{Khodjamirian:2010vf}, when using the same input parameters as in that paper.
    This difference becomes even larger when using up-to-date inputs, as explained in detail in the next subsection.
    We find that this reduction in size arises from cancellations across the Lorentz structures, since the ``new''
    structures enter the coefficient functions with opposite signs. Consequently, the phenomenological impact of the soft-gluon
    contribution to the non-local matrix elements is \emph{significantly reduced in the region where the LCOPE is applicable}.

\end{itemize}

\subsection{Numerical Results}
\label{sec:numres}

\begin{table}[t]
    \centering
    \renewcommand{\arraystretch}{1.2}
    \begin{tabular}{lclcclclc}
        \toprule
        Par.                             &
            Value                        &
            Units                        &
            Ref.                         &
            \hspace{1.0cm}               &
        Par.                             &
            Value                        &
            Units                        &
            Ref.                         \\
        \toprule
        $f_{B}$                          &
            $ 189.4 \pm 1.4$             &
            \MeV                         &
            \cite{Bazavov:2017lyh}       &
                                         &
        $f_{B_s}$                        &
            $ 230.7 \pm 1.3$             &
            \MeV                         &
            \cite{Bazavov:2017lyh}       \\
        $f_{K}$                          &
            $  155.6 \pm 0.4 $           &
            \MeV                         &
            \cite{Carrasco:2014poa}      &
                                         &
        $f_{K^*}$                        &
            $ 204 \pm 7 $                &
            \MeV                         &
            \cite{Straub:2015ica}        \\
        $f_{\phi}$                       &
            $ 233 \pm 4 $                &
            \MeV                         &
            \cite{Straub:2015ica}        &
                                         &
        $m_s(2\,\GeV)$ \hspace{-0.5cm}   &
            $95^{+9}_{-3}$               &
            \MeV                         &
            \cite{Tanabashi:2018oca}     \\
        $ \lambda_{B,+}$                 &
            $ \phantom{00} 460\pm 110$   &
            \MeV                         &
            \cite{Braun:2003wx}          &
                                         &
        $ \lambda_{B_s,+}$               &
            $ \phantom{00} 520\pm 110$   &
            \MeV                         &
            \cite{Bordone:2019guc}       \\
        $\lambda_{B_{(s)},E}^2$          &
            $\phantom{00}0.03\pm0.02$    &
            $\GeV^2$                     &
            \cite{Nishikawa:2011qk}      &
                                         &
        $\lambda_{B_{(s)},H}^2$          &
            $\phantom{00}0.06\pm0.03$    &
            $\GeV^2$                     &
            \cite{Nishikawa:2011qk}      \\
        $s_0^{B\to K}$                   &
            $1.05$                       &
            $\GeV^2$                     &
            \cite{Khodjamirian:2006st}   &
                                         &
        $s_0^{B\to K^*}$                 &
            $\phantom{0}[1.4, 1.7]$                 &
            $\GeV^2$                     &
            \cite{Gubernari:2018wyi}     \\
        $s_0^{B_s\to \phi}$              &
            $\phantom{0}[2.1, 2.4]$                 &
            $\GeV^2$                     &
            ---   \\
        \bottomrule
    \end{tabular}
    \renewcommand{\arraystretch}{1}
    \caption{ \it
        List of the input parameters used in the evaluation of the LCSR (\ref{eq:masterformula}).
        The intervals for the effective thresholds $s_0^{B\to K^*}$ and $s_0^{B_s\to \phi}$
        are the union of the $68\%$
        intervals of each individual form factor.
    }
    \label{tab:LCSRs:inputs}
\end{table}

The values of the parameters used in our numerical analysis are collected in \Tab{tab:LCSRs:inputs}.
For the $B \to K$ and $B \to K^*$ transitions,
these parameter coincides with the ones used in \Reff{Gubernari:2018wyi}.
In particular, we employ the same effective thresholds $s_0$ given in that paper.
This ensures consistency when using both local and non-local form factors in a simultaneous phenomenological analysis.

For the $B_s \to \phi$ transition, we need additional parameters that are not discussed in \Reff{Gubernari:2018wyi}.
We follow the procedure outlined in Ref.~\cite{Bordone:2019guc} to estimate the first inverse moment $1/\lambda_{B_s,+}$ of the leading twist $B_s$-LCDA. 
This estimate agrees with the recent calculation carried out in Ref.~\cite{Khodjamirian:2020hob}.
The values of the parameters $\lambda_{B_s,E}^2$ and $\lambda_{B_s,H}^2$, which enter in the models of $B_s$-LCDAs, are assumed to be equal to $\lambda_{B,E}^2$ and $\lambda_{B,H}^2$, respectively.
Given the large uncertainties of these latter parameters, we expect potential $SU(3)$-flavour symmetry-breaking effects to be negligible.
To determine the effective threshold for the LCSRs in the $B_s \to \phi$ transition, we first calculate the local $B_s \to \phi$ form factors using the analytical results of Ref.~\cite{Gubernari:2018wyi}.
In fact, these results can be employed to predict the local form factors for any $B \to V$ transition.
We can then set the effective threshold applying the procedure described in Ref.~\cite{Imsong:2014oqa}.
Our numerical predictions of the local $B_s \to \phi$ form factors are given in \App{app:BsPhiFFs}.
Note that this is the first calculation of these form factors using LCSRs with $B$-LCDAs.

For all the transitions considered, we vary the scale of the charm quark mass in the $\overline{\text{MS}}$ scheme between $m_c$ itself and $2 m_c$ and the Borel parameter in the interval $0.75\,\GeV^2 < M^2 < 1.25\,\GeV^2$, as in Ref.~\cite{Khodjamirian:2010vf}.
We verified that in this Borel window the tail of the LCOPE result  is much smaller than the LCOPE result integrated between 0 and $\sigma_0$.
Moreover, the sum rule dependence on $M^2$ is mild ($<6\%$ in the Borel window considered here) and negligible compared to the parametric uncertainties in our calculation.
\\

We can now evaluate the sum rule (\ref{eq:masterformula}) for $B \to K^{(*)}$ and $B_s \to \phi$ transitions.
The computer code needed to obtain our numerical results will be made publicly available under an open source
license as part of the \texttt{EOS} software~\cite{EOS}.
Our predictions for
$\tA[],\,\tV[]{1},\,\tV[]{2},$ and $\tV[]{3}$ are shown in \Tab{tab:LCSRscomp}.
For the $B\to K^{(*)}$ transitions we also compare our results with Ref.~\cite{Khodjamirian:2010vf}, while the results for the $B_s\to \phi$ transition are calculated for the first time.
One can easily observe that our results are roughly two orders of magnitude smaller than in Ref.~\cite{Khodjamirian:2010vf}.
As explained in the previous subsection, one order magnitude can be attributed to the different treatment of the three-particle $B$-LCDAs between the two papers.
The remaining difference is due to the updated input parameters used in our numerical analysis, in particular to the values of $\lambda_{B,E}^2$ and $\lambda_{B,H}^2$.
These parameters enter as the normalization of the three-particle $B$-LCDAs, and have therefore a large impact on the overall size of the hadronic matrix elements.
In Ref.~\cite{Khodjamirian:2010vf} the approximation $\lambda_{B,E}^2 = \lambda_{B,H}^2$ is adopted, which is not justified by calculations of these parameters~\cite{Descotes-Genon:2019bud,Grozin:1996pq,Nishikawa:2011qk}:
\begin{align}
    \frac{\lambda_{B,E}^2}{\lambda_{B,H}^2}
    =
    0.4^{+0.5}_{-0.3}
    \,.
\end{align}
In Ref.~\cite{Khodjamirian:2010vf} it is also assumed that $\lambda_{B,E}^2=\frac{3}{2}\lambda_{B,+}^2$, based solely on the desire that the
exponential model for the leading-twist $B$-LCDA satisfies exactly the Grozin-Neubert relations~\cite{Grozin:1996pq}. This assumption yields a central
value for $\lambda_{B,E}^2$ approximately $20$ times bigger than the one found in its most recent calculation~\cite{Nishikawa:2011qk}.
We do not use this rather strong assumption, and use the calculated values instead.

We emphasize that although the relative uncertainties of our results listed in \Tab{tab:LCSRscomp} are similar to the ones in Ref.~\cite{Khodjamirian:2010vf}, our absolute uncertainties are much smaller.
\\

\begin{table}[t!]
    \newcommand{\pp}{\phantom{+}}
    \centering
    \setlength{\tabcolsep}{12pt}
    \renewcommand{\arraystretch}{1.45}
    \begin{tabular}{cccc}
        \toprule
        Transition & $\tV[]{}(q^2 = 1\,\GeV^2)$ & This work & Ref.~\cite{Khodjamirian:2010vf} 
        \\\toprule
        $B \to K$ & 
        $\tA[]$ & 
        $ (+4.9\pm2.8)\cdot 10^{-7}\phantom{\, \GeV}$ &
        $ (-1.3^{+1.0}_{-0.7})\cdot 10^{-4}\phantom{\, \GeV}$  
        \\\midrule
                   &
        $\tV[]{1}$ & 
        $(-4.4\pm3.6)\cdot 10^{-7} \, \GeV$ &
        $(-1.5^{+1.5}_{-2.5})\cdot 10^{-4} \, \GeV$ 
        \\
        $B \to K^*$ & 
        $\tV[]{2}$ & 
        $ (+3.3\pm2.0)\cdot 10^{-7} \, \GeV$ &
        $(+ 7.3^{+14}_{-7.9})\cdot 10^{-5} \, \GeV$  
        \\
                   &
        $\tV[]{3}$ &  
        $(+1.1\pm1.0)\cdot 10^{-6} \, \GeV$ &
        $(+ 2.4^{+5.6}_{-2.7})\cdot 10^{-4} \, \GeV$ 
        \\\midrule
                   &
        $\tV[]{1}$ & 
        $(-4.4\pm5.6)\cdot 10^{-7} \, \GeV$ &
        --- 
        \\
        $B_s \to \phi$ & 
        $\tV[]{2}$ & 
        $ (+4.3\pm3.1)\cdot 10^{-7} \, \GeV$ &
        --- 
        \\
                   &
        $\tV[]{3}$ &  
        $(+1.7\pm2.0)\cdot 10^{-6} \, \GeV$ &
        --- 
        \\\bottomrule
    \end{tabular}
    \caption{ \it
    \label{tab:LCSRscomp} 
      Comparison between the results of Ref.~\cite{Khodjamirian:2010vf} and our results at $q^2 = 1\,\GeV^2$.
    }
\end{table}

\begin{table}
    \newcommand{\pp}{\phantom{+}}
    \centering
    \renewcommand{\arraystretch}{1.6}
    \resizebox{1.0\textwidth}{!}{%
    \begin{tabular}{cccccccccc}
        \toprule 
        Transition                       &
        Pol.                             &
        $q^2 \,[\GeV^2]$                 &
        $\FM{\lambda}$                   &
        $\FM{\lambda,T}$                 &
        $\tV{\lambda}$                   &
        $\Re\HM{\lambda}$                &
        $\Im\HM{\lambda}$                \\
        \toprule
        \multirow{4}{*}{\rotatebox{90}{$B \to K$}}       &
        \multirow{4}{*}{$\lambda=0$}     &
        $-7$                             &
        $+0.191\pm0.055$                 &
        $-0.041\pm0.010$                 &
        $(-8.0\pm3.8)\cdot10^{-8}$       &
        $(+3.8\pm7.3)\cdot10^{-6}$       &
        $(-7.5\pm1.9)\cdot10^{-7}$       \\
                                         &
                                         &
        $-5$                             &
        $+0.209\pm0.061$                 &
        $-0.032\pm0.008$                 &
        $(-5.1\pm2.4)\cdot10^{-8}$       &
        $(+6.1\pm6.8)\cdot10^{-6}$       &
        $(-7.4\pm1.9)\cdot10^{-7}$       \\
                                         &
                                         &
        $-3$                             &
        $+0.230\pm0.067$                 &
        $-0.021\pm0.005$                 &
        $(-2.2\pm1.1)\cdot10^{-8}$       &
        $(+6.4\pm5.4)\cdot10^{-6}$       &
        $(-6.5\pm2.4)\cdot10^{-7}$       \\
                                         &
                                         &
        $-1$                             &
        $+0.254\pm0.074$                 &
        $-0.008\pm0.002$                 &
        $(-2.2\pm1.5)\cdot10^{-9}$       &
        $(+3.4\pm2.4)\cdot10^{-6}$       &
        $(-3.2\pm0.9)\cdot10^{-7}$       \\
        \midrule
        \multirow{12}{*}{\rotatebox{90}{$B \to K^*$}}    &
        \multirow{4}{*}{$\lambda=\perp$} &
        $-7$                             &
        $+0.354\pm0.117$                 &
        $+0.365\pm0.117$                 &
        $(-0.5\pm5.1)\cdot10^{-7}$       &
        $(+1.3\pm0.4)\cdot10^{-4}$       &
        $(+6.6\pm2.1)\cdot10^{-6}$       \\
                                         &
                                         &
        $-5$                             &
        $+0.363\pm0.120$                 &
        $+0.376\pm0.120$                 &
        $(-0.4\pm4.7)\cdot10^{-7}$       &
        $(+1.4\pm0.4)\cdot10^{-4}$       &
        $(+8.0\pm2.6)\cdot10^{-6}$       \\
                                         &
                                         &
        $-3$                             &
        $+0.373\pm0.119$                 &
        $+0.385\pm0.123$                 &
        $(-0.4\pm3.8)\cdot10^{-7}$       &
        $(+1.4\pm0.4)\cdot10^{-4}$       &
        $(+9.9\pm3.3)\cdot10^{-6}$       \\
                                         &
                                         &
        $-1$                             &
        $+0.382\pm0.122$                 &
        $+0.394\pm0.126$                 &
        $(-0.4\pm2.5)\cdot10^{-7}$       &
        $(+1.4\pm0.4)\cdot10^{-4}$       &
        $(+1.2\pm0.4)\cdot10^{-5}$       \\
        \cline{2-8}
                                         &
        \multirow{4}{*}{$\lambda=\para$} &
        $-7$                             &
        $+0.382\pm0.122$                 &
        $+0.357\pm0.114$                 &
        $(-2.1\pm5.2)\cdot10^{-7}$       &
        $(+1.3\pm0.4)\cdot10^{-4}$       &
        $(+6.5\pm2.0)\cdot10^{-6}$       \\
                                         &
                                         &
        $-5$                             &
        $+0.397\pm0.123$                 &
        $+0.368\pm0.118$                 &
        $(-2.0\pm4.7)\cdot10^{-7}$       &
        $(+1.4\pm0.4)\cdot10^{-4}$       &
        $(+7.9\pm2.5)\cdot10^{-6}$       \\
                                         &
                                         &
        $-3$                             &
        $+0.410\pm0.127$                 &
        $+0.381\pm0.122$                 &
        $(-1.8\pm3.9)\cdot10^{-7}$       &
        $(+1.4\pm0.4)\cdot10^{-4}$       &
        $(+9.8\pm3.2)\cdot10^{-6}$       \\
                                         &
                                         &
        $-1$                             &
        $+0.425\pm0.131$                 &
        $+0.393\pm0.130$                 &
        $(-1.5\pm2.6)\cdot10^{-7}$       &
        $(+1.4\pm0.4)\cdot10^{-4}$       &
        $(+1.2\pm0.4)\cdot10^{-5}$       \\
        \cline{2-8}
                                         &
        \multirow{4}{*}{$\lambda=0$}     &
        $-7$                             &
        $+0.256\pm0.171$                 &
        $-0.074\pm0.047$                 &
        $(-1.6\pm0.7)\cdot10^{-7}$       &
        $(+0.0\pm2.7)\cdot10^{-5}$       &
        $(-1.3\pm0.9)\cdot10^{-6}$       \\
                                         &
                                         &
        $-5$                             &
        $+0.271\pm0.181$                 &
        $-0.055\pm0.035$                 &
        $(-1.1\pm0.5)\cdot10^{-7}$       &
        $(+0.4\pm2.4)\cdot10^{-5}$       &
        $(-1.2\pm0.8)\cdot10^{-6}$       \\
                                         &
                                         &
        $-3$                             &
        $+0.279\pm0.187$                 &
        $-0.034\pm0.022$                 &
        $(-5.8\pm2.7)\cdot10^{-8}$       &
        $(+0.5\pm1.7)\cdot10^{-5}$       &
        $(-1.0\pm0.6)\cdot10^{-6}$       \\
                                         &
                                         &
        $-1$                             &
        $+0.292\pm0.199$                 &
        $-0.012\pm0.008$                 &
        $(-1.5\pm0.8)\cdot10^{-8}$       &
        $(+3.0\pm7.2)\cdot10^{-6}$       &
        $(-4.7\pm3.1)\cdot10^{-7}$       \\
        \midrule
        \multirow{12}{*}{\rotatebox{90}{$B_s \to \phi$}} &
        \multirow{4}{*}{$\lambda=\perp$} &
        $-7$                             &
        $+0.417\pm0.112$                 &
        $+0.470\pm0.127$                 &
        $(-0.2\pm5.6)\cdot10^{-7}$       &
        $(+1.6\pm0.4)\cdot10^{-4}$       &
        $(+8.4\pm2.3)\cdot10^{-6}$       \\
                                         &
                                         &
        $-5$                             &
        $+0.426\pm0.115$                 &
        $+0.456\pm0.123$                 &
        $(-0.2\pm5.1)\cdot10^{-7}$       &
        $(+1.6\pm0.4)\cdot10^{-4}$       &
        $(+9.6\pm2.6)\cdot10^{-6}$       \\
                                         &
                                         &
        $-3$                             &
        $+0.433\pm0.121$                 &
        $+0.441\pm0.119$                 &
        $(-0.1\pm4.3)\cdot10^{-7}$       &
        $(+1.6\pm0.4)\cdot10^{-4}$       &
        $(+1.1\pm0.3)\cdot10^{-5}$       \\
                                         &
                                         &
        $-1$                             &
        $+0.440\pm0.123$                 &
        $+0.423\pm0.114$                 &
        $(-0.2\pm2.9)\cdot10^{-7}$       &
        $(+1.5\pm0.4)\cdot10^{-4}$       &
        $(+1.3\pm0.4)\cdot10^{-5}$       \\
        \cline{2-8}
                                         &
        \multirow{4}{*}{$\lambda=\para$} &
        $-7$                             &
        $+0.458\pm0.119$                 &
        $+0.294\pm0.080$                 &
        $(-2.4\pm5.8)\cdot10^{-7}$       &
        $(+1.2\pm0.3)\cdot10^{-4}$       &
        $(+5.3\pm1.4)\cdot10^{-6}$       \\
                                         &
                                         &
        $-5$                             &
        $+0.472\pm0.123$                 &
        $+0.314\pm0.085$                 &
        $(-2.3\pm5.3)\cdot10^{-7}$       &
        $(+1.2\pm0.3)\cdot10^{-4}$       &
        $(+6.6\pm1.8)\cdot10^{-6}$       \\
                                         &
                                         &
        $-3$                             &
        $+0.488\pm0.127$                 &
        $+0.335\pm0.087$                 &
        $(-2.1\pm4.4)\cdot10^{-7}$       &
        $(+1.3\pm0.3)\cdot10^{-4}$       &
        $(+8.4\pm2.3)\cdot10^{-6}$       \\
                                         &
                                         &
        $-1$                             &
        $+0.503\pm0.131$                 &
        $+0.357\pm0.093$                 &
        $(-1.8\pm3.1)\cdot10^{-7}$       &
        $(+1.3\pm0.3)\cdot10^{-4}$       &
        $(+1.1\pm0.3)\cdot10^{-5}$       \\
        \cline{2-8}                      &
        \multirow{4}{*}{$\lambda=0$}     &
        $-7$                             &
        $+0.315\pm0.180$                 &
        $-0.088\pm0.035$                 &
        $(-1.7\pm0.7)\cdot10^{-7}$       &
        $(+0.0\pm2.7)\cdot10^{-5}$       &
        $(-1.6\pm0.9)\cdot10^{-6}$       \\
                                         &
                                         &
        $-5$                             &
        $+0.330\pm0.188$                 &
        $-0.066\pm0.035$                 &
        $(-1.1\pm0.5)\cdot10^{-7}$       &
        $(+0.5\pm2.3)\cdot10^{-5}$       &
        $(-1.5\pm0.8)\cdot10^{-6}$       \\
                                         &
                                         &
        $-3$                             &
        $+0.341\pm0.194$                 &
        $-0.041\pm0.021$                 &
        $(-5.8\pm2.7)\cdot10^{-8}$       &
        $(+0.6\pm1.7)\cdot10^{-5}$       &
        $(-1.2\pm0.6)\cdot10^{-6}$       \\
                                         &
                                         &
        $-1$                             &
        $+0.354\pm0.206$                 &
        $-0.014\pm0.007$                 &
        $(-1.5\pm0.8)\cdot10^{-8}$       &
        $(+3.5\pm7.0)\cdot10^{-6}$       &
        $(-5.5\pm7.0)\cdot10^{-7}$       \\
        \bottomrule
    \end{tabular}
    }
    \caption{ \it
    \label{tab:RESULTS} 
      Our predictions for the quantities $\FM{\lambda,(T)}$, $\tV{\lambda}$, and $\HM{\lambda}$ for different values of $q^2$.
      See text for details.
    }
\end{table}

Our predictions for $\FM{\lambda,(T)}$, $\tV{\lambda}$, and $\HM{\lambda}$ at $q^2 = \lbrace -7, -5, -3, -1\rbrace\,\GeV^2$ are shown in \Tab{tab:RESULTS}.
The values of the local form factors $\FM{\lambda,(T)}$ are taken from the LCSRs calculation of Ref.~\cite{Gubernari:2018wyi}\,\footnote{
        The uncertainties of the local form factors can be further reduced by using combined fits to lattice QCD and LCSR results~\cite{Straub:2015ica,Gubernari:2018wyi}.
}.
The non-local form factors $\HM{\lambda}$ are computed using \Eq{eq:masterNLFF}.
The numerical results for $\Delta C_7$ and $\Delta C_9$, needed to evaluate $\HM{\lambda}$, are obtained from the \texttt{Mathematica} notebook attached to the arXiv version of~Ref.~\cite{Asatrian:2019kbk}:
\begin{align*}
    \Delta C_7 (q^2 = -7\,\GeV^2) = & -0.05571 -0.00362 \,i \,,&
    \Delta C_9 (q^2 = -7\,\GeV^2) = &\,0.10228 +0.00001 \,i \,,\\
    \Delta C_7 (q^2 = -5\,\GeV^2) = & -0.05850 -0.00431 \,i \,,&
    \Delta C_9 (q^2 = -5\,\GeV^2) = &\,0.13181 -0.00039 \,i \,,\\
    \Delta C_7 (q^2 = -3\,\GeV^2) = & -0.06150 -0.00519 \,i \,,&
    \Delta C_9 (q^2 = -3\,\GeV^2) = &\,0.16603 -0.00114 \,i \,,\\
    \Delta C_7 (q^2 = -1\,\GeV^2) = & -0.06472 -0.00637 \,i \,,&
    \Delta C_9 (q^2 = -1\,\GeV^2) = &\,0.20715 -0.00263 \,i \,.
\end{align*}
In anticipation of the next section, we keep only the contributions proportional to the
charm quark electric charge $Q_c$ in the above results.
Their uncertainties are negligible compared to the ones of the local form factors.

The findings of Ref.~\cite{Khodjamirian:2010vf} imply that 
the next-to-leading power contribution in the LCOPE could be larger than the leading power contribution in the computation of the non-local form factors $\HM{\lambda}$.
This has been cause for concern about the rate of convergence of the LCOPE even at spacelike momentum transfer.
One of the main findings of our work is that the contribution at next-to-leading power is in fact negligible compared to the theory uncertainties of the leading-power term. The authors of Ref.~\cite{Khodjamirian:2010vf} come to a different conclusion, due to the missing terms in the calculation of the hadronic matrix element.
As a consequence, theoretical predictions of the $\HM{\lambda}$ are dominated by the leading-power of the LCOPE, i.e. stem
from the first two terms in \Eq{eq:masterNLFF}.
Therefore, our findings thoroughly eliminate the concern and give confidence that a precise theoretical prediction in the spacelike
region is now possible.


\section{Dispersive Bound}
\label{sec:bounds}
\setcounter{equation}{0}

The results for the non-local form factors $\HM{\lambda}$ of~\Sec{sec:subleading-operators} need to be analytically continued from the spacelike
region of $q^2$, where they are obtained, to the timelike region, where they are required for phenomenological studies. 
This requires a suitable parametrization of the hadronic matrix elements.
Previous parametrizations based on series expansions involve an uncontrollable
truncation error~\cite{Khodjamirian:2010vf,Ciuchini:2015qxb,Jager:2012uw,Bobeth:2017vxj}.
In the case of local form factors, this problem is solved by imposing dispersive bounds, which provide control over the systematic truncation
errors by turning them into parametric errors.
However, no such bound has been derived for non-local matrix elements as of yet.\\

The purpose of this section is to derive the dispersive bound for the non-local matrix elements for the first time.
To this end, we construct a parametrization of these matrix elements that manifestly satisfies a dispersion relation
derived from the total cross section of $e^+e^- \to b\bar{s} X$.
We obtain the dispersive bound by matching two representations of the discontinuity due to $b\bar{s}$ on-shell states
of a suitable correlation function $\Pi$:
\begin{equation}
\begin{aligned}
\label{eq:Pi-general}
\Pi^{\mu\nu}(q)
\equiv
i \int  d^4 x\, e^{i q\cdot x}\bra{0} T\left\lbrace O^\mu(q;x), O^{\nu,\dagger}(q;0) \right\rbrace  \ket{0} 
=
\left(\frac{q^\mu q^\nu}{q^2} - g^{\mu\nu}\right) \Pi(q^2)\,.
\end{aligned}
\end{equation}
Here, the operators $O^\mu(q;x)$ and $O^{\dagger,\nu}(q;0)$ are defined as\,\footnote{The notation is such that $O^\mu(q;0) = -\frac{1}{q^2} \K^\mu(q)$ of~\Reff{Asatrian:2019kbk}.}
\begin{equation}
\begin{aligned}
    \label{eq:non-loc-oper}
    O^\mu(q;x)
        & = \left(\frac{-16\pi^2 i}{q^2}\right)\int d^4y\, e^{+i q \cdot y}\ 
            T\big\{ j_\text{em}^\mu(x+y), (C_1 \cO_1 + C_2 \cO_2)(x)\big\}
            \,,\\
    O^{\nu,\dagger}(q;0)
        & = \left(\frac{+16\pi^2 i}{q^2}\right)\int d^4z\, e^{-i q\cdot z}\ 
            T\big\{ j_\text{em}^\nu(z), (C_1 \cO_1 + C_2 \cO_2)^\dagger(0)\big\}\,.
\end{aligned}
\end{equation}
The invariant correlation function $\Pi$ has two classes of contributions to its discontinuity $\text{Disc}\ \Pi = \Disc \Pi + \Disccc \Pi$ : $\Disc \Pi$ arising from intermediate flavored on-shell states
with strangeness and beauty $B = -S = -1$; and $\Disccc \Pi$ (for our discussion irrelevant) arising from intermediate unflavored on-shell states with $B = -S = 0$.
Since the strong and electromagnetic interactions conserve flavor, the separation of these two types of discontinuities is well defined to all orders in $\alpha_s$ and $\alpha_e$.
The non-local matrix elements $\HM{\mu}$ contribute to $\Disc \Pi$.

The discontinuity $\Disc \Pi$ satisfies a subtracted dispersion relation:
\begin{align}
    \label{eq:def-chi}
    \chi(Q^2)
        \equiv \frac{1}{n!}\left[\frac{d}{dQ^2}\right]^n \frac{1}{2 i\pi} \int\limits_{0}^\infty ds\ \frac{\Disc \Pi(s)}{s-Q^2}
    \,.
\end{align}
Here $n$ is the yet-to-be-determined number of subtractions and $Q^2$ is the subtraction point, chosen so that an OPE can be performed.
We first isolate the contribution to $\Pi$ that stems exclusively from the $b\bar{s}$ cut in \refsec{bounds:bsbar-cut},
thereby determining $\Disc \Pi$ using a local OPE.
We then derive the hadronic dispersion relation for $\chi$ in terms of the non-local hadronic matrix elements in \refsec{bounds:HDR}.
We then use our knowledge from the previous two sections to construct a parametrization of the non-local hadronic matrix elements
that manifestly fulfils a dispersive bound on its parameters in \refsec{bounds:derivation}.
We finally present a practical application of the dispersive bound in \refsec{application}.

\subsection{Calculation of the Discontinuity in a Local OPE}
\label{sec:bounds:bsbar-cut}

We now calculate the contributions to the discontinuity of the correlation function~(\ref{eq:Pi-general}) that exclusively arise from $b\bar s$-flavoured intermediate states.
To simplify this task, we use the low-recoil OPE for the operators in~\refeq{non-loc-oper} well within its region of applicability, that is for $q^2 \geq (m_b + m_s)^2$~\cite{Grinstein:2004vb,Beylich:2011aq}:
\begin{equation}
    \label{eq:high-q2-OPE}
    O^\mu(q;x) = \sum_{d,n} C_{d,n}(q^2) \,O_{d,n}^\mu(q;x)\ .\\
\end{equation}
Here, $d$ is the mass dimension of the local operators $O^{\mu}_{d,n}$, while $n$ labels the different operators with the same mass dimension. The Wilson coefficients of operators of dimension $d$ scale as $(\Lambda_\text{had} / m_b)^{d-3}\sim (\Lambda_\text{had} / \sqrt{q^2})^{d-3} \sim 0.1^{d-3}$.
The first few operators in this expansion read \cite{Beylich:2011aq}
\begin{equation}
\begin{aligned}
    O_{3,1}^\mu(q,x)   & = \left(g^{\mu\nu} - \frac{q^\mu q^\nu}{q^2}\right) [\bar{s} \gamma_\nu P_L b](x)\,, &&
    \hspace{-25mm} O_{3,2}^\mu(q,x)    = -\frac{2i m_b}{q^2}  q_\nu [\bar{s} \sigma^{\mu\nu} P_R b](x)\,, \\[2mm]
    O_{4,1}^\mu(q,x)   & = m_s O_{3,1}^\mu(q,x)\,, &&
   \hspace{-25mm}  O_{4,2}^\mu(q,x)    = m_s O_{3,2}^\mu(q,x)\,,\\[2mm]
    O_{5,1}^\mu(q,x) &= \frac{1}{q^2} \left(\eps^{\alpha q \lambda \rho} q^\mu + \eps^{q \mu \lambda \rho} q^\alpha - \eps^{\alpha\mu\lambda\rho} q^2\right)
                    [\bar{s} \gamma_\lambda G_{\alpha\rho} P_R b](x)\,.
\end{aligned}
\end{equation}
The Wilson coefficients of the leading dimension-three operators are given by
\begin{equation}
\begin{aligned}
    C_{3,1}(q^2) & = f^{(9)}_{\text{LO}}(q^2) - \frac{\alpha_s}{4 \pi} \left[C_1 F_{1c}^{(9)}(q^2) + C_2 F_{2c}^{(9)}(q^2)\right]\,, \\
    C_{3,2}(q^2) & = -\frac{\alpha_s}{4 \pi} \left[C_1 F_{1c}^{(7)}(q^2) + C_2 F_{2c}^{(7)}(q^2)\right]\,,
\end{aligned}
\end{equation}
where we use the same definition of $f^{(9)}_{\text{LO}}$ and $F_{1c,2c}^{(7,9)}$ as in~\Reff{Asatrian:2019kbk},
thereby only retaining the contributions proportional to the charm quark electric charge $Q_c$.
Here and in this section we use $\alpha_s \equiv \alpha_s(m_b)$.
Already in the calculation of $C_{3,1}$ at leading order in $\alpha_s$ one encounters UV divergences in dimensional regularization.
Hence,
the coefficients $C_{3,j}$ are always understood to be renormalized~\cite{Asatryan:2001zw}.

A few comments are in order regarding the OPE:
\begin{itemize}
    \item The Wilson coefficients of the operators of dimensions 4 and 5 start at $\order{\alpha_s}$;
    \item operators of dimensions $d=3,4$ interfere with each other, but these interference terms
    arise only at order $\alpha_s\,m_s / m_b$;
    \item the operators at dimension $d \geq 5$ do not interfere with the ones at mass dimension $d=3$ or $d=4$
    to leading-order in $\alpha_s$.
\end{itemize}
Based on the above, we adopt the power counting $\eps^2 \sim \Lambda_\text{had}/m_b \sim \alpha_s^2$.
Thus, up to corrections of order $\eps^3$, we can express the discontinuity of $\Pi$ as
\begin{equation}
    \label{eq:Pi:OPE}
    \begin{aligned}
    \Disc \Pi^\OPE (s)
        & = \left|C_{3,1}(s)\right|^2 \, \Disc \Pi_{1,1}(s) + \left|C_{3,2}(s)\right|^2  \, \Disc \Pi_{2,2}(s)\\
        & + 2  \Re\left\lbrace C_{3,1}(s) C^{*}_{3,2}(s)\right\rbrace \, \Disc \Pi_{1,2}(s) + \order{\eps^3}\ ,
    \end{aligned}
\end{equation}
where we have used the short hand notation
\begin{align}
    \Pi_{i,j}(s)
        & \equiv
         \frac{1}{D-1}\left(\frac{q_\mu q_\nu}{q^2} - g_{\mu\nu}\right)
         i\int d^4 x\, e^{i q\cdot x}\,
            \bra{0} T\left\lbrace
               O_{3,i}^{\mu} (q,x),
               O_{3,j}^{\nu,\dagger}(q,0)
            \right\rbrace \ket{0}
            \ .
\end{align}
It is instructive to investigate the perturbative expansion of $\Disc \Pi^\OPE$ by writing $\Pi_{i,j}(s)=\Pi_{i,j}^\text{LO}+\Pi_{i,j}^\text{NLO}+ \order{\alpha_s^2}$ and $C_{i,j}(s)=C_{i,j}^\text{LO}+C_{i,j}^\text{NLO}+ \order{\alpha_s^2}$:
\begin{equation}
\begin{aligned}
    \Disc\Pi^\OPE(s) =
        & \left|C^\LO_{3,1}\right|^2 \Disc\Pi_{1,1}^\LO + \left|C^\LO_{3,1}\right|^2 \Disc\Pi_{1,1}^\NLO\\
        & \hspace{-20mm} + 2 \Re\left\{C^\LO_{3,1}C^{\NLO,*}_{3,1}\right\} \Disc\Pi_{1,1}^\LO
         + 2  \Re\left\lbrace C^\text{LO}_{3,1} C^{\NLO,*}_{3,2}\right\rbrace \, \Disc\Pi_{1,2}^\LO
          + \order{\eps^2}\ .
\end{aligned}
\end{equation}
To \LO we find the compact expressions
\begin{align}
     \Disc\Pi_{1,1}^\LO(s)
    & =
    -\frac{i  \lamkin ^{1/2}
   \left(\lamkin +3 s \left(m_b^2+m_s^2-s\right)\right)}{8 \pi
    s^2}
    \theta \left(s-(m_b+m_s)^2\right)\ ,
   \\
     \Disc\Pi_{2,2}^\LO(s)
    & =
     -\frac{i m_b^2  \lamkin
   ^{1/2} \left(2 \lamkin +3 s
   \left(m_b^2+m_s^2-s\right)\right)}{2 \pi  s^3}
   \theta \left(s-(m_b+m_s)^2\right)\ ,
   \\
     \Disc\Pi_{1,2}^\LO(s)
    & =
    -\frac{3 i m_b m_s \left(m_b^2-m_s^2+s\right)  \lamkin ^{1/2}}{4 \pi  s^2}
    \theta\left(s-(m_b+m_s)^2\right)\ .
\end{align}
Here $\lamkin \equiv \lambda(m_b^2,m_s^2,s)$.
The \NLO expressions have been calculated in the context of the gauge boson self-energies~\cite{Djouadi:1993ss}.
The one needed here is given by
\begin{align}
    \Disc\Pi_{1,1}^\NLO(s) =  i s  \frac{\alpha_s}{\pi}\frac1{4\pi^2} \Im \Pi_T^+(s)\,,
\end{align}
where $\Im \Pi_T^+(s)$ has been calculated in~\Reff{Djouadi:1993ss}.
Inserting these results in \refeq{def-chi} we find that at least two subtractions ($n = 2$) are needed in \refeq{def-chi},
and thus
\begin{align}
    \label{eq:chi-result}
    \chi^\OPE(Q^2) \equiv
    \frac{1}{2 i\pi}
    \int\limits_{0}^\infty
     ds\  \frac{\Disc \Pi^\OPE(s)}{(s-Q^2)^3}
    \,.
\end{align}
For $Q^2 = \lbrace 0, -m_b^2\rbrace$ we obtain
\begin{equation}
\begin{aligned}
    \chi^\OPE(-m_b^2) & = (1.81\pm0.02)\cdot 10^{-4}\ \text{GeV}^{-2}\,, \\
    \chi^\OPE(0) & = (2.69\pm0.03)\cdot 10^{-4}\ \text{GeV}^{-2}\,,
\end{aligned}
\end{equation}
where the quoted uncertainties are only due to varying $m_b=4.18_{-0.02}^{+0.03}\,\text{GeV}$~\cite{Zyla:2020zbs}.

\subsection{Hadronic Dispersion Relation}
\label{sec:bounds:HDR}

By means of unitarity, the discontinuity of the correlation function $\Pi$ can be expressed in terms of
a sum of sesquilinear combinations of hadronic matrix elements:
\eq{
\Disc \Pi \sim \sum_n \av{0|O^\mu|n} \av{n|O^\dagger_\mu|0}\ ,
}
with $\ket{n}$ labelling all possible on-shell states. The matrix elements with $|n\rangle = |M\bar B\rangle$ are related by crossing symmetry  to the non-local matrix elements $\HM{\mu}$ of \Eq{eq:def-Hmu}.
Since the $T$-product in~\Eq{eq:Pi-general} is a Hermitian operator, all the contributions to $\Disc \Pi$ must be positive definite, and so
one can find an upper bound on the non-local form factors $\HM{\lambda}$ by
ignoring the contributions from all other states. Including additional states in a simultaneous analysis
would further strengthen the bound. 
\\

The one-body contributions to $\Disc \Pi$ involve $\bar{B}_s^*$-to-vacuum matrix elements of the non-local operators.
While we do not include these contributions here, they can be easily accounted for in future works.
The two-body contributions to $\Disc \Pi$ arise from intermediate $\bar{B}K$, $\bar{B}K^*$, $\bar{B_s}\phi$ and
further $b\bar{s}$ states that also include baryons, such as $\Lambda_b\bar\Lambda$. Their contributions
to $\Disc \Pi$ can be expressed as follows:
\begin{align}           
    \label{eq:ImPi-hadronic0}
    \left(\frac{q^\mu q^\nu}{q^2} - g^{\mu\nu}\right) 
    \Disc\Pi^\text{had}(s)
        & = 
            i
            \sum \!\!\!\!\!\!\!\!\!\! \int \limits_{H_bH_{\bar{s}}} d\rho_{H_b H_{\bar{s}}}\, (2\pi)^4 \delta^{(4)}(p_{H_b H_{\bar{s}}} - q)\\*
    \nonumber
        & \times \,
            \braket{0| O^\mu (q; 0)| H_bH_{\bar{s}}}
            \braket{H_bH_{\bar{s}} | O^{\dagger,\nu}(q; 0) | 0}\\*
    \nonumber
        & + \text{further positive terms}\,.
\end{align}
Here $H_b$ and $H_{\bar{s}}$ denote hadrons with flavour quantum numbers $B=-1$ and $S=1$, respectively,
and the two-body phase space measure is given by
\eq{
\int d\rho_{X\, Y} (2\pi)^4 \delta^{(4)}(p_{X\,Y} - q)
= \frac{1}{8\pi}\,\frac{\sqrt{\lambda(M_X^2, M_Y^2, s)}}{s}\, \theta(s - s_{XY})\ ,
\label{eq:PSmeasure}
}
with $s_{XY}\equiv (M_X + M_Y)^2$.
Since we work in the isospin limit, the contributions due to $\bar{B}^0K^{(*)0}$ and $B^- K^{(*)+}$ are identical.
Hence, we simply multiply the $\bar{B}^0K^{(*)0} = B^- K^{(*)+} \equiv \bar{B}K^{(*)}$ contributions by a factor of $2$.
Keeping only the contributions due to $\bar{B}K^{(*)}$ and $\bar{B}_s \phi$, and using \Eq{eq:def:correlator-BKstar}, we find
\eqa{
&& \hspace{-3mm}  \frac{3}{32 i\pi^3}\, \Disc \Pi^\text{had}(s)
        = \frac{2 M_B^4\,\lambda^{3/2}(M_B^2, M_K^2, s)}{ s^4} \left|\HP[K]{0}(s)\right|^2
            \theta(s - s_{BK})
\nonumber\\
&&        +\ \frac{2 M_B^6\,\sqrt{\lambda(M_{B_{}}^2, M_{K^*}^2, s)}}{ s^3} \left(
            \left|\HV[B\to K^*]{\perp}(s)\right|^2
            + \left|\HV[B\to K^*]{\para}(s)\right|^2
            + \frac{M_B^2}{s}\left|\HV[B\to K^*]{0}(s)\right|^2
        \right) \theta(s - s_{BK^*})
\nonumber\\
&&        +\ \frac{M_B^6\,\sqrt{\lambda(M_{B_s}^2, M_{\phi}^2, s)}}{ s^3} \left(
            \left|\HV[B_s\to\phi]{\perp}(s)\right|^2
            + \left|\HV[B_s\to\phi]{\para}(s)\right|^2
            + \frac{M_{B_s}^2}{s} \left|\HV[B_s\to\phi]{0}(s)\right|^2
        \right) \theta(s - s_{B_s\phi})
\nonumber\\[2mm]
&&        +\ \text{further positive terms .}
\label{eq:ImPi-hadronic}
}
The non-local matrix elements $\HM{\mu}$ develop a series of branch cuts starting at $q^2 = 4M_D^2$, that is below the $(M_B + M_M)^2$ threshold.
Although they do not contribute to $\Disc \Pi$, they still spoil the analyticity of the non-local form factors $\HM{\lambda}$
in the semileptonic region $0 \leq q^2 \leq (M_B - M_M)^2$, which is the phenomenologically interesting one.
This makes the derivation of the bound for non-local matrix elements considerably more complicated than the one for local matrix elements (see e.g.~\Reff{Bharucha:2010im}).
In particular, 
it implies that the coefficients of the Taylor expansion in the variable $z$ of the non-local form factors do not fulfil a dispersive bound.
In the next section, we show that appropriately chosen
functions of $z$, which fulfil a non-trivial orthogonality relation on the integration domain, cure this problem.

\subsection{Derivation of the Bound}
\label{sec:bounds:derivation}

We start by matching the OPE result onto the hadronic representations of $\chi(Q^2)$ --- defined in \refeq{def-chi} with $n=2$ --- by means of global quark-hadron duality:
\begin{align}
    \label{eq:sche-bounds}
    \frac{1}{2 \pi i} \int\limits_{0}^\infty  ds\,\frac{\Disc \Pi^\OPE(s)}{(s-Q^2)^{3}} = \frac{1}{2 i\pi} \int\limits_{0}^\infty ds\, \frac{\Disc \Pi^\text{had}(s)}{(s-Q^2)^{3}}
    \ .
\end{align}
We then use \refeq{ImPi-hadronic} to rewrite \refeq{sche-bounds} as a dispersive bound on weighted integrals of the hadronic matrix elements:
\begin{align}
\chi^\OPE(Q^2)
&= \frac{32 \pi^2}{3} \int\limits_{(M_B + M_{K})^2}^{\infty} ds\, 
                \frac{M_B^4\,\lambda^{3/2}(M_{B}^2, M_K^2, s)}{ s^4 (s-Q^2)^3}\, \left|\HP[K]{0}(s)\right|^2
\nonumber\\
&\hspace{-4mm} + \frac{32 \pi^2}{3} \int\limits_{(M_B + M_{K^*})^2}^{\infty} ds\,
                \frac{M_B^6\,\sqrt{\lambda(M_{B_{}}^2, M_{K^*}^2, s)}}{ s^3(s-Q^2)^3} 
                \left(\sum_{\lambda=\perp,\para} \left|\HV[B\to K^*]{\lambda}(s)\right|^2
                +\frac{M_B^2}{s}\left|\HV[B\to K^*]{0}(s)\right|^2
                \right)
\nonumber\\
&\hspace{-4mm}+ \frac{16 \pi^2}{3} \int\limits_{(M_{B_s} + M_{\phi})^2}^{\infty} ds\,
        \frac{M_{B_s}^6\,\sqrt{\lambda(M_{B_s}^2, M_{\phi}^2, s)}}{ s^3(s-Q^2)^3} 
        \left(\sum_{\lambda=\perp,\para} \left|\HV[B_s\to\phi]{\lambda}(s)\right|^2
        +\frac{M_{B_s}^2}{s}\left|\HV[B_s\to \phi]{0}(s)\right|^2
        \right)
\nonumber\\*[2mm]
&\hspace{-4mm} + \text{further positive terms .}
\end{align}
Following the usual procedure to obtain dispersive bounds of the parameters of the hadronic matrix elements~\cite{Boyd:1995sq}, we define the map
\begin{equation}
    \label{eq:zdef}
    z(s) \equiv \frac{\sqrt{s_+ - s} - \sqrt{s_+ - s_0 }}{\sqrt{s_+ - s} + \sqrt{s_+ - s_0 }}\,.
\end{equation}
Here $s_+$ is the lowest branch point of the matrix element and $s_0$ can be chosen freely in the open interval  $(-\infty,\,s_+)$. In our case, we have $s_+ = 4 M_D^2$ rather than $(M_B + M_{K^{(*)}})^2$ as
in the case for the local $B\to K^{(*)}$ form factors.
Using this map and the fact that $z = e^{i\alpha}$ on the unit circle, we obtain
\begin{align}
\chi^\OPE(Q^2)
& = \frac{16 \pi^2}{3} \int\limits_{-\alpha_{BK}}^{+\alpha_{BK}} 
\left.
d\alpha\, \left|\frac{dz(\alpha)}{d\alpha} \frac{ds(z)}{dz}\right| \frac{M_B^4\,\lambda^{3/2}(M_B^2, M_K^2, s)}{ s^4 (s-Q^2)^3}\, \left|\HP[K]{0}(s)\right|^2\right|_{\scriptsize \begin{array}{c}s=s(z)\\[-1mm] z=z(\alpha)\end{array}}
\nonumber\\
&+  \frac{16 \pi^2}{3}  \int\limits_{-\alpha_{BK^*}}^{+\alpha_{BK^*}} d\alpha\, \left|\frac{dz(\alpha)}{d\alpha} \frac{ds(z)}{dz}\right|\frac{M_B^6\,\sqrt{\lambda(M_{B_{}}^2, M_{K^*}^2, s)}}{ s^3(s-Q^2)^3} \left.
     \sum_{\lambda=\perp,\para} \left|\HV[B\to K^*]{\lambda}(s)\right|^2
                \right|_{\scriptsize \begin{array}{c}s=s(z)\\[-1mm] z=z(\alpha)\end{array}}
\nonumber\\
&+  \frac{16 \pi^2}{3}  \int\limits_{-\alpha_{BK^*}}^{+\alpha_{BK^*}} 
\left.
d\alpha\, \left|\frac{dz(\alpha)}{d\alpha} \frac{ds(z)}{dz}\right|\frac{M_B^8\,\sqrt{\lambda(M_{B_{}}^2, M_{K^*}^2, s)}}{ s^4(s-Q^2)^3} 
      \left|\HV[B\to K^*]{0}(s)\right|^2
                \right|_{\scriptsize \begin{array}{c}s=s(z)\\[-1mm] z=z(\alpha)\end{array}}
\nonumber\\
& + \frac{8 \pi^2}{3} \int\limits_{-\alpha_{B_s\phi}}^{+\alpha_{B_s\phi}} d\alpha\, \left|\frac{dz(\alpha)}{d\alpha} \frac{ds(z)}{dz}\right|\frac{M_{B_s}^6\,\sqrt{\lambda(M_{B_s}^2, M_{\phi}^2, s)}}{ s^3(s-Q^2)^3}  \left.
\sum_{\lambda=\perp,\para} \left|\HV[B_s\to\phi]{\lambda}(s)\right|^2
        \right|_{\scriptsize \begin{array}{c}s=s(z)\\[-1mm] z=z(\alpha)\end{array}}
        \nonumber\\
& + \frac{8 \pi^2}{3} \int\limits_{-\alpha_{B_s\phi}}^{+\alpha_{B_s\phi}}
\left.
d\alpha\, \left|\frac{dz(\alpha)}{d\alpha} \frac{ds(z)}{dz}\right|\frac{M_{B_s}^8\,\sqrt{\lambda(M_{B_s}^2, M_{\phi}^2, s)}}{ s^4(s-Q^2)^3}  
\left|\HV[B_s\to \phi]{0}(s)\right|^2
        \right|_{\scriptsize \begin{array}{c}s=s(z)\\[-1mm] z=z(\alpha)\end{array}}
\nonumber\\*
& + \text{further positive terms ,}
\label{eq:disp-rel-z}
\end{align}
where the integral limits are given by
\begin{equation}
    \alpha_{XY} \equiv \big|\arg z(s_{XY})\big|\ ,
\end{equation}
with $s_{XY}$ defined previously below~\Eq{eq:PSmeasure}.

The central improvement of this paper is the change of the parametrization discussed in~\Reff{Bobeth:2017vxj}
to one that fulfils a dispersive bound.
As in that paper, we remove the dynamical singularities of the non-local form factors $\HM{\lambda}$
using the \emph{Blaschke factor} 
\begin{equation}
\P(z) \equiv \prod_{\psi=J/\psi,\psi(2S)} \frac{z - z_\psi}{1 - z \, z_\psi^*}\,.
\end{equation}
Here $z_\psi = z(s = M_\psi^2)$ is the location of the two narrow charmonium poles in the complex $z$ plane.
These are the only poles on the open unit disk.
In addition, to formulate the bound in a concise form and to avoid kinematical singularities, we 
introduce suitable \emph{outer functions} $\outerF{\lambda}(z)$~\cite{Rudin}.
These  outer functions are defined such that on the integration domain their
modulus squared coincides with \Eqs{eq:weightBK}{eq:weightBsphi} \emph{and} they are free
of unphysical singularities inside the unit disk. 
The precise form of these functions and their derivation is provided in~\App{app:fs}.
We can then define the functions
\eqa{
\label{eq:HhatBK}
\hHP[P]{0}(z) & \equiv & \outerF[B\to P]{0}(z) \, \P(z) \,\HP[P]{0}(z)\,,
\\*
\label{eq:HhatBV}
\hHV[B\to V]{\lambda}(z) & \equiv & \outerF[B\to V]{\lambda}(z) \, \P(z) \, \HV[B\to V]{\lambda}(z) \,,
}
which are analytical on the open unit disk.
\\

At this point, we can express the dispersive bound as
\begin{multline}
\label{eq:dispbound}
1 > 2 \!\!\int\limits_{-\alpha_{BK}}^{+\alpha_{BK}} \!\!d\alpha \left|\hHP[K]{0}(e^{i\alpha})\right|^2
+
\sum_{\lambda}
\left[
2\!\!
\int\limits_{-\alpha_{BK^*}}^{+\alpha_{BK^*}} \!\!d\alpha
 \left|\hHV[B\to K^*]{\lambda}(e^{i\alpha}) \right|^2
    +\!\! \int\limits_{-\alpha_{B_s\phi}}^{+\alpha_{B_s\phi}} \!\!d\alpha
    \left|\hHV[B_s\to \phi]{\lambda}(e^{i\alpha}) \right|^2\, 
    \right].
\end{multline}
The next step is to find a basis of orthogonal functions
on an arc of the unit circle covering angles $-\alpha_{XY}$ to $+\alpha_{XY}$. In lieu of a closed formula, we construct the first
three orthonormal polynomials $p_{n}^{X\to Y}$ as
\begin{align}
\label{eq:poly}
    p_{0}^{X\to Y}(z)
        & = \frac{1}{\sqrt{2 \alpha_{XY}}}\,,\\
    \nonumber
    p_{1}^{X\to Y}(z)
        & = \left(z - \frac{\sin(\alpha_{XY})}{\alpha_{XY}} \right) \sqrt{\frac{\alpha_{XY}}{2\alpha_{XY}^2 + \cos(2\alpha_{XY}) - 1}}\,, \\
    \nonumber
    p_{2}^{X\to Y}(z)
        & = \left(z^2 + \frac{\sin(\alpha_{XY})(\sin(2\alpha_{XY}) - 2\alpha_{XY})}{2\alpha_{XY}^2 + \cos(2\alpha_{XY}) - 1} z + \frac{2\sin(\alpha_{XY}) (\sin(\alpha_{XY}) - \alpha_{XY}\cos(\alpha_{XY}))}{2\alpha_{XY}^2 + \cos(2\alpha_{XY}) - 1}\right)\\
    \nonumber
        & \hspace{-8mm} \times \sqrt{\frac{
        2(2\alpha_{XY}^2 + \cos(2\alpha_{XY}) - 1)
        }{
        -9\alpha_{XY}+8\alpha_{XY}^3+8\alpha_{XY} \cos(2\alpha_{XY})+\alpha_{XY} \cos(4\alpha_{XY})+4 \sin(2\alpha_{XY}) - 2 \sin(4\alpha_{XY})
        }}\,.
\end{align}
The higher order polynomials can be determined using an orthogonalization procedure.
For an in-depth review of the mathematical properties of these orthogonal polynomials on the unit circle, we refer to~\Reff{Simon2005}.
The practical considerations of this application of the orthogonal polynomials are discussed in~\Reff{Eberhard2020}.

Using the orthogonal polynomials, we can now expand
\begin{equation}
    \label{eq:expansionHhat}
    \hHM{\lambda}(z)
        = \sum_{n=0}^\infty  a_{\lambda,n}^{B\to M} p_{n}^{B\to M}(z)\,.
\end{equation}
The dispersive bound then takes the simple form
\begin{equation}
    \label{eq:boundcoeff}
     \sum_{n=0}^\infty 
    \left\{
        2\Big| a_{0,n}^{B\to K} \Big|^2
        +
        \sum_{\lambda=\perp,\para,0}
        \left[
            2\Big| a_{\lambda,n}^{B\to K^*} \Big|^2
            +
            \Big| a_{\lambda,n}^{B_s\to \phi} \Big|^2
    \right]
    \right\}
     < 1\,.
\end{equation}
It is worth noting that the coefficients of the Taylor expansions of the local form factors --- multiplied by suitable outer functions and Blaschke factors --- satisfy an analogous constraint.
In that case, the $z$ monomials constitute a complete and orthonormal basis of polynomials on the integration domain, which is the unit circle in the $z$ plane.
We have seen that the integration domain for the non-local form factor case is only an arc of the unit circle, due to the appearance of $D\bar{D}$ and similar branch cuts
below the $\bar{B}M$ thresholds.
As a consequence, the orthonormal polynomials in this integration domain are the ones given in \Eq{eq:poly}.
While these polynomials are clearly much more complicated than the $z$ monomials, they allow us to write the dispersive bound in the diagonal form shown in (\ref{eq:boundcoeff}).

An inconvenient feature of the $p_{n}^{B\to M}$ polynomials is that their magnitude increases for $n\to \infty$ in the semileptonic region.
Nevertheless, since the series in \Eq{eq:expansionHhat} is convergent for $|z|<1$ due to the analyticity of $\hHM{\lambda}$, this only implies that the coefficients $a_{\lambda,n}^{B\to M}$ must fall off sufficiently fast such that higher order terms in the series are suppressed.

In the next subsection we present a simple application of the bound in~\Eq{eq:boundcoeff} to $\HP[K]{0}$.
We remark that it is also possible to expand the $\hHM{\lambda}$ functions in terms of $z$ monomials using the same Blaschke factors and outer functions give here. 
However, the coefficients of that expansion do \emph{not} satisfy any dispersive bound.

\subsection{Application to $\bar B \to \bar K \ell^+\ell^-$}
\label{sec:application}

\begin{figure}
    \centering
    \includegraphics[width=8cm]{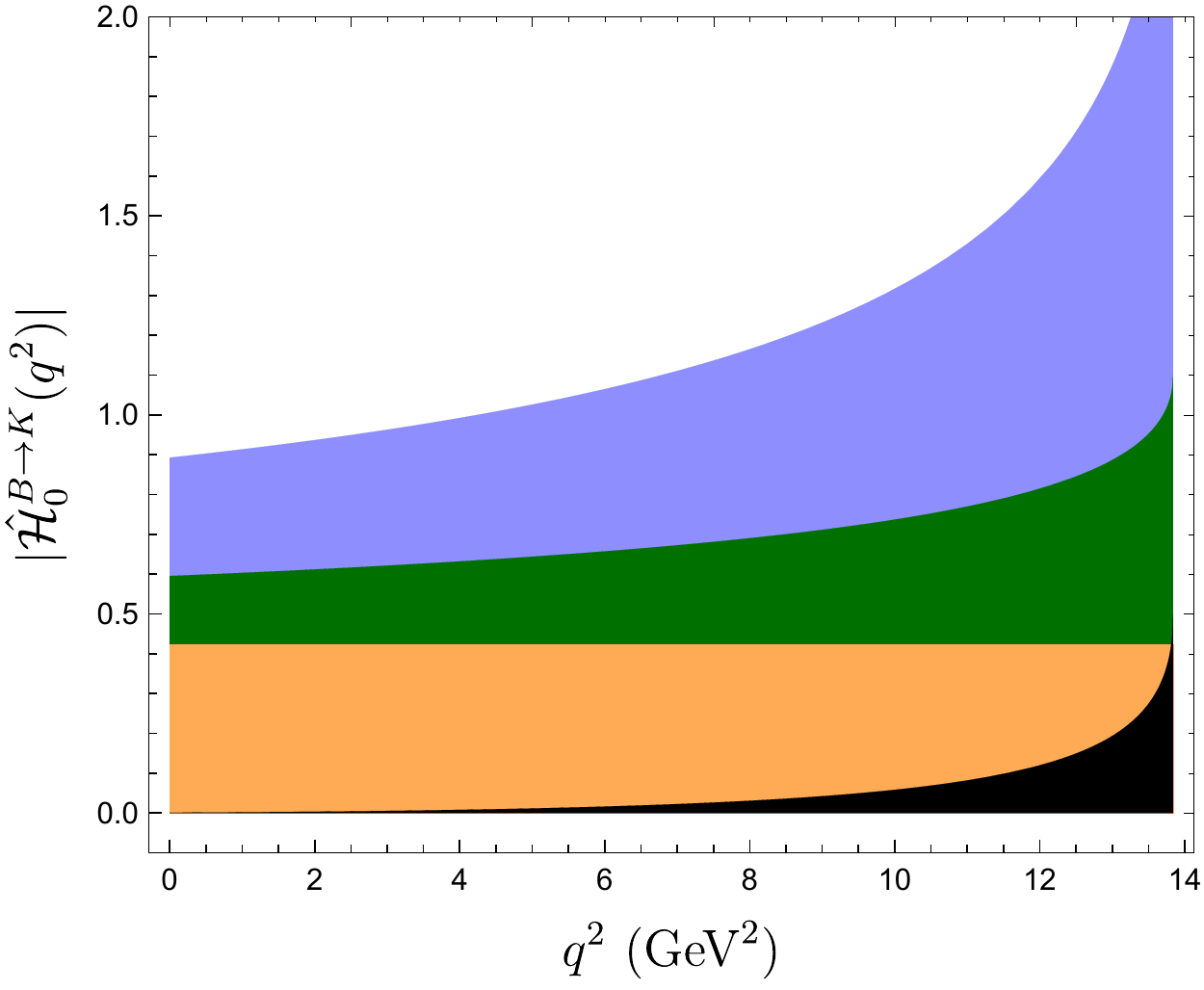}
    \hspace{4mm}
    \includegraphics[width=8cm]{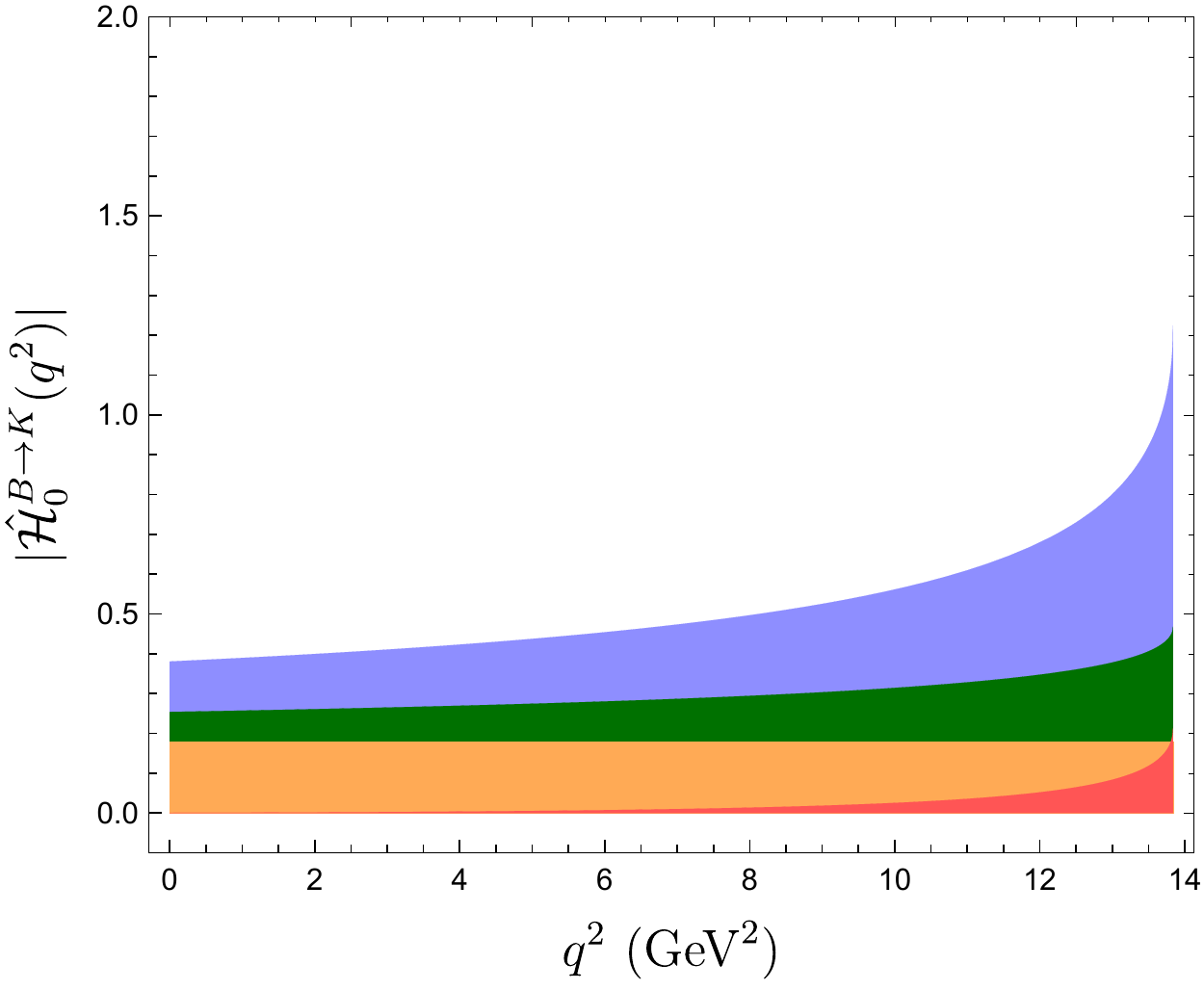}
    \caption{\it Left: Allowed values for the magnitude of $\hHP[K]{0}$ according to the dispersive bound on the expansion with one (orange), two (green) and three (blue) coefficients. The black region shows the allowed region for the three-coefficient expansion including two theory constraints at $q^2=-1\,\GeV$ and $-5\,\GeV$.
    Right: The same, assuming all $\bar B\to \bar K^{(*)}\ell^+\ell^-$ and $\bar B_s\to \phi \ell^+\ell^-$ modes contribute equally to the bound. In this case the region that includes the theory constraints is shown in red.}
    \label{fig:Hhat}
\end{figure}

We now explore some of the implications of the dispersive bound (\ref{eq:boundcoeff}). 
Considering only the $\bar B\to \bar K\ell^+\ell^-$ contribution to the bound, we have
\eq{
\hHP[K]{0}(z)= \sum_{n=0}^N a_{n}^{B\to K} p_n^{B\to K}(z)
\, , \qquad
\sum_{n=0}^N \left|a_{n}^{B\to K}\right|^2 < \frac12\ ,
}
where we have assumed that the series expansion is truncated at $n=N$.
Depending on the value of $N$, the bound sets a global constraint on the size of $|\hHP[K]{0}|$. This is shown in the left panel of~\Fig{fig:Hhat}, where the constraints for $N=0,1,2$ are shown in yellow, green and blue, respectively. 
In order to express the result as a function of $q^2$, we have taken $s_0=0$.

\begin{figure}
    \centering
    \includegraphics[scale=1.25]{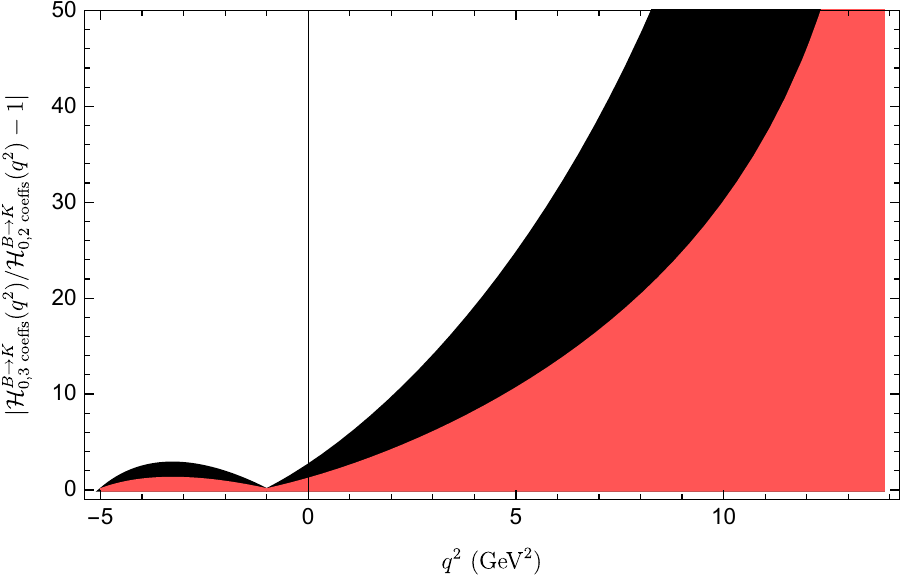}
    \caption{\it Estimate of the (relative) truncation error in the determination of $\HP[K]{0}$ with two coefficients and two theory constraints at $q^2=-1\,\GeV$ and $-5\,\GeV$. The black region disregards the contribution from other modes (most conservative situation), while the red region assumes equal contributions from all $\bar B\to \bar K^{(*)}\ell^+\ell^-$ and $\bar B_s\to \phi \ell^+\ell^-$ modes (c.f.~\Fig{fig:Hhat}).}
    \label{fig:HKratio}
\end{figure}

Of course, one may include the theoretical calculation based on the LCOPE at negative $q^2$. 
In order to see how this information impacts the constraints on the size of $|\hHP[K]{0}|$, we impose that
$\HP[K]{0}$ takes the central values quoted in~\Tab{tab:RESULTS} at $q^2=-1\,\GeV^2$ and $q^2=-5\,\GeV^2$. 
At this point we need to make a choice on the value of the subtraction point $Q^2$.
Here we take $Q^2=-m_b^2$ in the outer functions.
In the case $N=2$, these two theory constraints fix two independent (complex-valued) combinations of $a_{0,1,2}^{B\to K}$, leaving one complex free parameter. 
This free parameter is then constrained by the dispersive bound, and leads to the black region shown in the left panel of~\Fig{fig:Hhat}.
This black region could be regarded as an estimate of the truncation error when using two theory data points at negative $q^2$ to fix the $N=1$ series expansion of $\hHP[K]{0}(q^2)$. In relative terms, one may consider the ratio of the resulting allowed values for $\hHP[K]{0}(q^2)$ with $N=2$ to the theoretical curve for the $N=1$ expansion fixed by the two theory data points. This is shown by the black region in~\Fig{fig:HKratio}.

This example does not take full power of the dispersive bound, in particular by neglecting the fact that other modes (e.g. $\bar B\to \bar K^*\ell^+\ell^-$ and $\bar B_s\to \phi\ell^+\ell^-$) also contribute to the bound. 
An analysis that takes this into account will lead to simultaneously correlated bounds for all $\HP[K]{0}(q^2)$,  $\HP[K^*]{\lambda}(q^2)$ and $\HV[B_s\to \phi]{\lambda}(q^2)$, and thus it is beyond the scope of this section. In order to estimate what the impact of adding these other modes might be in a simple setting, we can make the reasonable simplifying assumption that all eleven modes in~\Eq{eq:boundcoeff} contribute equally to the bound, and thus take
\eq{
\sum_{n=0}^N \left|a_{n}^{B\to K}\right|^2 < \frac1{11}\qquad \text{(involves\ assumption)} 
\ .
}
The corresponding bounds are shown in the right panel of~\Fig{fig:Hhat}, and the red region in~\Fig{fig:HKratio}. In this case the constraint is tightened by more than a factor of two.

\begin{figure}
    \centering
    \includegraphics[width=8cm]{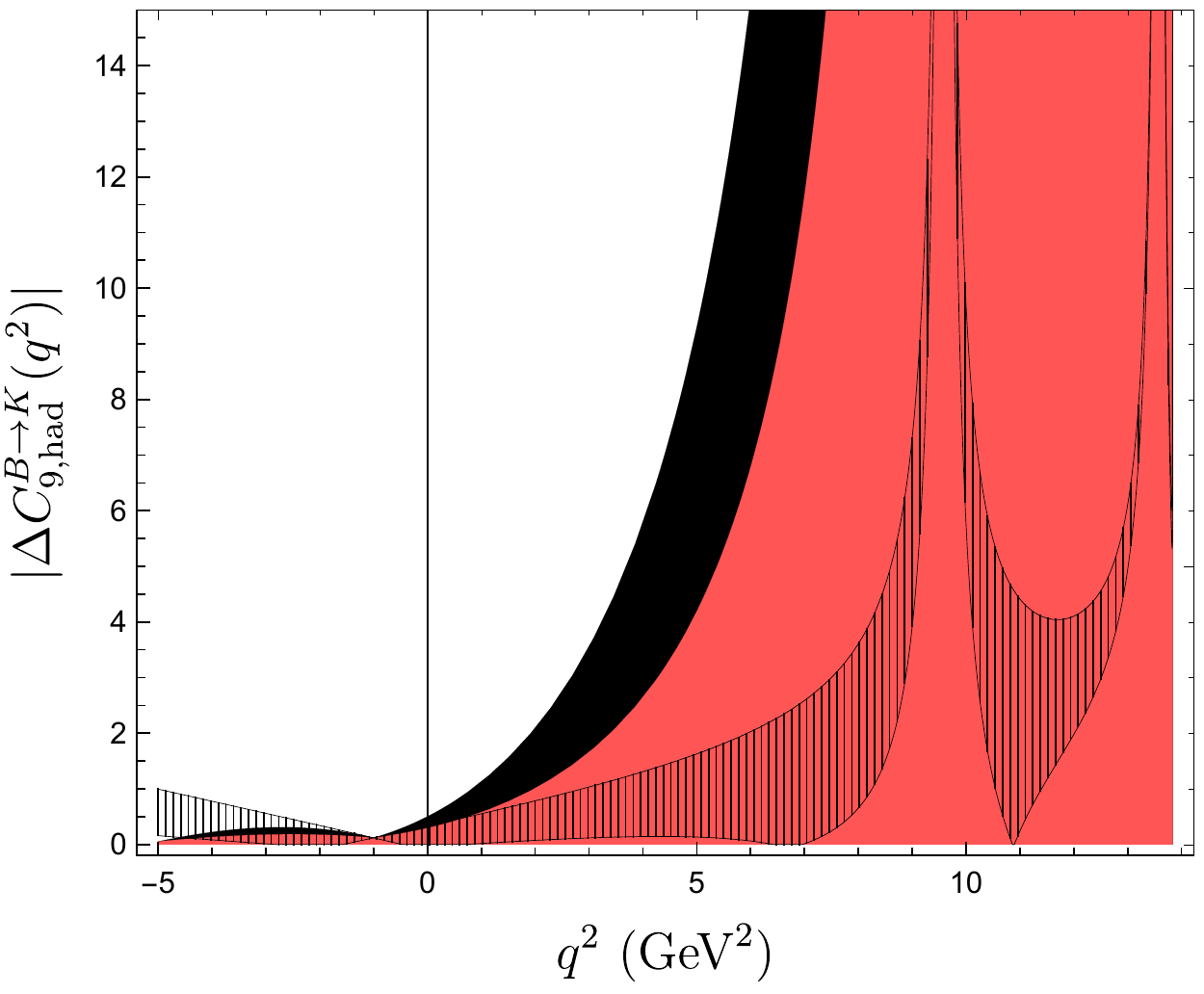}
    \hspace{4mm}
    \includegraphics[width=7.9cm]{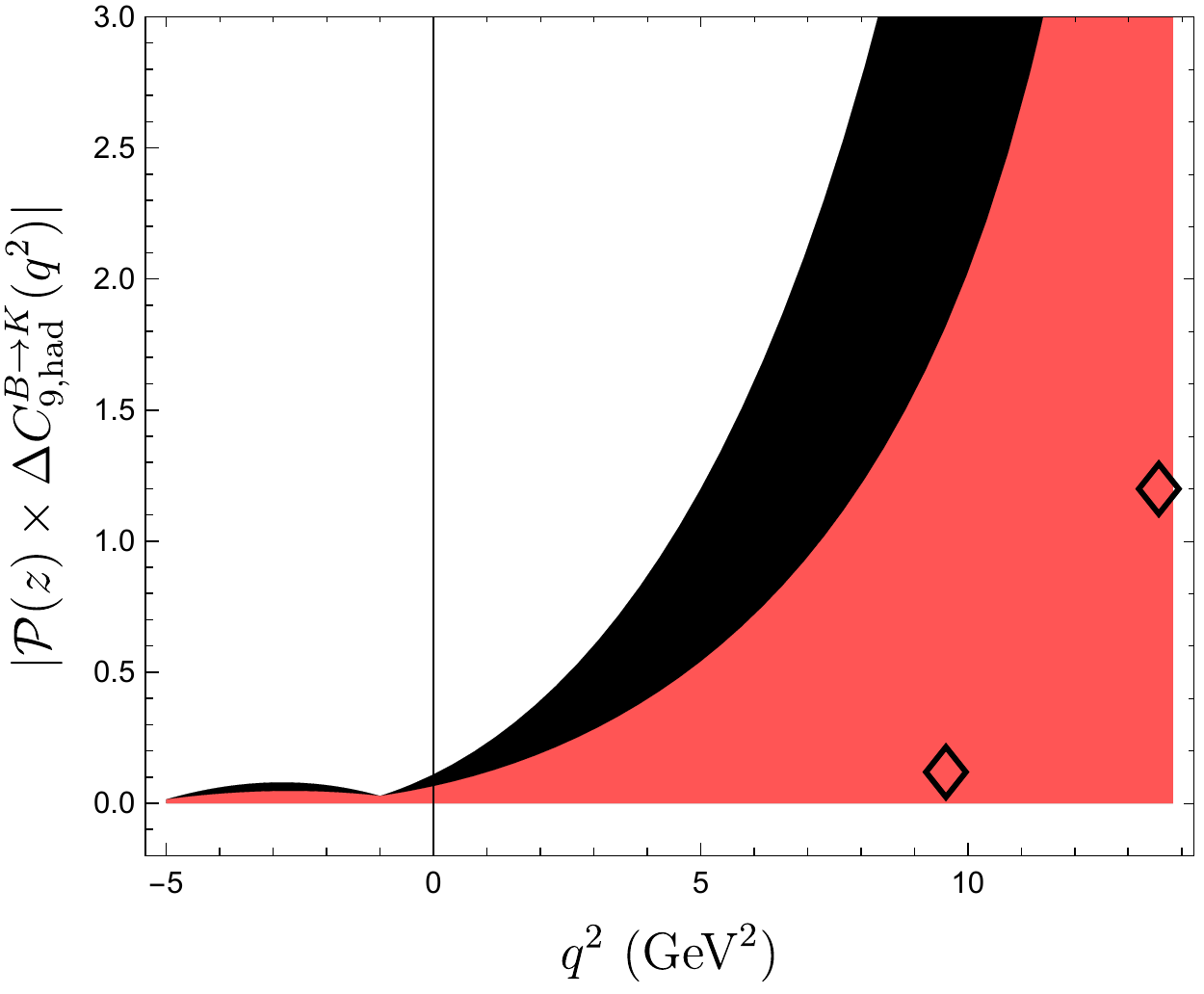}
    \caption{\it Allowed values for the magnitudes of $\Delta C^{B\to K}_{9,\text{had}}(q^2)$ and the combination $\P(z) \Delta C^{B\to K}_{9,\text{had}}(q^2)$.
    Same color coding as in~\Fig{fig:HKratio}. The diamonds in the right-hand plot show the experimental data points on the
    charmonium poles (central values only). The resulting constraints on $|\Delta C^{B\to K}_{9,\text{had}}(q^2)|$ are shown as the shaded region in the left
    plot (including only one theory data point).  See text for details.}
    \label{fig:DeltaC9}
\end{figure}

To finish, it is interesting to see what these bounds look like at the amplitude level in comparison to the contribution from $C_9$. To that end, we consider the quantity
\eq{
\Delta C^{B\to K}_{9,\text{had}}(q^2) = \frac{32\pi^2 M_B^2}{q^2} \frac{\HP[K]{0}(q^2)}{\FP[K]{0}(q^2)}\ ,
}
which is defined in such a way that
\eq{
\A(\bar B\to M \ell^+\ell^-) = \frac{G_F\, \alpha\, V^*_{ts} V_{tb}}{\sqrt{2} \pi}  (C_9 -  \Delta C^{B\to K}_{9,\text{had}})  \,L^\mu_{V}\ \FP[K]{\mu}  + \cdots  \,.
}
The corresponding limits on the magnitude of $\Delta C^{B\to K}_{9,\text{had}}(q^2)$ in the same circumstances as those in~\Fig{fig:HKratio} are shown \Fig{fig:DeltaC9} (left panel). One can see that the non-local contribution cannot exceed (in magnitude) the SM value for $C_9(m_b) \simeq 4$ for $q^2\lesssim 3\,\GeV^2$, or even for $q^2\lesssim 5\,\GeV^2$ in the simplified scenario where the $K^*$ and $\phi$ modes are included.

The raise of the bound for $q^2\to 9\,\GeV^2$ is due to the fact that $\Delta C^{B\to K}_{9,\text{had}}(q^2)$ contains a pole at $q^2=M_{J/\psi}^2$. In this sense it may be convenient to consider instead the combination ${\cal P}(z) \times \Delta C^{B\to K}_{9,\text{had}}(q^2)$ where the narrow charmonium poles have been removed. This is shown in the right panel of \Fig{fig:DeltaC9}. This quantity is also directly related with the $B\to K \psi_n$ amplitudes, at $q^2=M^2_{\psi_n}$, and thus the experimental measurement of these amplitudes can be used to constrain further the non-local form factor~\cite{Khodjamirian:2010vf,Khodjamirian:2012rm,Bobeth:2017vxj}. In particular, with the conventions of~\Reff{Bobeth:2017vxj},
\eq{
\operatorname*{Res}_{q^2\to M_{\psi_n}^2}
\Delta C^{B\to K}_{9,\text{had}}(q^2) = \frac{16\pi^2 f_{\psi_n}^* \A_{\psi_n}}{\FP[K]{0}(M_{\psi_n}^2)}\ .}
Taking into account that
\eq{
\lim_{q^2\to M_{\psi_n}^2} \frac{{\cal P}(z)}{q^2-M_{\psi_n}^2} = \left\{
\begin{array}{rl}
-0.036\,\GeV^{-2} & \quad\text{for } J/\psi \\
0.604\,\GeV^{-2} & \quad\text{for } \psi(2S)
\end{array}
\right.
}
one finds
\eqa{
|{\cal P}(z_{J/\psi})\,\Delta C^{B\to K}_{9,\text{had}}(M_{J\psi}^2)| &=&
(0.036\,\GeV^{-2})\, \frac{16\pi^2 f_{J/\psi} |\A_{J/\psi}|}{\FP[K]{0}(M_{J/\psi}^2)} \simeq 0.12\ ,
\\
|{\cal P}(z_{\psi(2S)})\,\Delta C^{B\to K}_{9,\text{had}}(M_{\psi(2S)}^2)| &=&
(0.604\,\GeV^{-2})\, \frac{16\pi^2 f_{\psi(2S)} |\A_{\psi(2S)}|}{\FP[K]{0}(M_{\psi(2S)}^2)}  \simeq  1.2\ ,
}
where we have used
\begin{align}
f_{J/\psi} &\simeq  0.277\,\GeV\ ,
&\A_{J/\psi} &\simeq 0.035\,\GeV \ ,
&\FP[K]{0}(M_{J/\psi}^2) &\simeq 0.47\ ,
\\
f_{\psi(2S)} &\simeq 0.198\,\GeV\ ,
&\A_{\psi(2S)} &\simeq 0.041\,\GeV \ ,
&\FP[K]{0}(M_{\psi(2S)}^2)  &\simeq 0.65\ .
\end{align}
These experimental data points are shown in the right-hand plot of~\Fig{fig:DeltaC9}, and are situated well within the dispersive bounds, as required.
Including this experimental information in the determination of $\HP[K]{0}(q^2)$ is rather important~\cite{Khodjamirian:2010vf,Khodjamirian:2012rm,Bobeth:2017vxj}, since it essentially turns the \emph{extrapolation} from the spacelike to the timelike region into an \emph{interpolation}, up to undetermined strong phases that are not fixed
by the non-leptonic amplitudes. The result of adding these two charmonium data points on the allowed ranges for $|\Delta C^{B\to K}_{9,\text{had}}(q^2)|$ is shown by the dashed region in the left-hand plot of~\Fig{fig:DeltaC9}. In this case, only the theory data point at $q^2=-1\,\GeV$ has been kept, since otherwise the system would be overconstrained.
Again, the resulting region is situated well within the dispersive bound.

A more detailed and complete phenomenological study of the implications of the dispersive bound and the determination of the non-local contributions to $\bar B\to M\ell\ell$ amplitudes is left for future work.

\section{Summary and Conclusions}
\label{sec:conclusion}

The contributions from four-quark effective operators are an essential part of the exclusive
$\bar{B}_{(s)}\to \lbrace \bar{K}^{(*)}, \phi\rbrace\ell^+\ell^-$ and $\bar{B}_{(s)}\to \lbrace \bar{K}^{*}, \phi\rbrace \gamma$
amplitudes. These contributions must be under reasonable theoretical control in order to derive solid conclusions from the measurements of such decay observables. However, they enter the decay amplitudes through a non-local matrix element of non-perturbative nature,
which is very difficult to calculate with controlled uncertainties. In this work we have revisited this non-local effect and made
progress at two fronts.

First, we have recalculated the main subleading effect beyond the local OPE contribution, which arises from a soft gluon coupling to a
quark loop. This recalculation involves a light-cone sum rule with $B$-meson light-cone distribution amplitudes ($B$-LCDAs), and improves upon the only
previous calculation of this quantity by including the full set of $B$-LCDAs up to and including twist-four.
Our reanalysis leads to a result for this soft-gluon effect that is two orders of magnitude smaller than the previous calculation.
A substantial part of the difference is due to cancellations arising from the inclusion of the $B$-LCDAs that were missing. The remaining difference is due to updated inputs.

Second, we have revisited the analytic continuation of the non-local effect from the LCOPE region (where it is calculated) to the
physical region relevant for $B$ decays. In particular, we have proposed a modified analytic parametrization of the non-local matrix
element and derived a dispersive bound that constrains this parametrization. 
The combination of our new parametrization with the dispersive bound allows for the first time to control the inevitable
systematic truncation error from which every existing parametrization of the matrix elements suffers.

Our results lead to a better understanding of the non-local contributions to decay modes such as
$\bar B\to \bar K^{(*)}\ell^+\ell^-$ and $\bar B_s\to \phi\ell^+\ell^-$, which are currently under intense experimental and theoretical
scrutiny. An important question is whether the anomalies observed in these modes are due to physics beyond the SM.
Our ability to answer this question hinges on our ability to bound poorly known QCD effects that could potentially
be responsible for the discrepancies.
We can now say that the first subleading correction to the hadronic non-local contribution is very small in the decays considered here,
giving support to theory calculations that neglect this subleading effect. In addition, our dispersive bound sets a solid ground for the
analytic continuation of calculations from the LCOPE region to the physical one. Future phenomenological applications of the results
presented here will lead to more accurate global analyses of $b\to s\ell^+\ell^-$ data.

\section*{Acknowledgements}

We are very grateful to Alexander Khodjamirian, Sebastian Jäger, Roman Zwicky and Yu-Ming Wang for helpful discussions. We also would
like to acknowledge helpful discussion with Christoph Bobeth, who participated in the first steps of this project.

The work of NG and DvD is supported by the Deutsche Forschungsgemeinschaft (DFG, German Research Foundation)
within the Emmy Noether Programme under grant DY-130/1-1 and the DFG Collaborative Research Center 110
``Symmetries and the Emergence of Structure in QCD''.
The research of NG is supported by the Deutsche Forschungsgemeinschaft (DFG, German Research Foundation) under grant  396021762 - TRR 257.
JV acknowledges funding from the European Union's Horizon 2020 research and innovation programme under the Marie Sklodowska-Curie grant agreement No 700525
`NIOBE', and from the Spanish MINECO through the ``Ram\'on y Cajal'' program RYC-2017-21870.
The authors would like to express a special thanks to the Mainz Institute for Theoretical Physics (MITP) of the Cluster of Excellence PRISMA+
(Project ID 39083149) for its hospitality and support.

\appendix
\addtocontents{toc}{\protect\setcounter{tocdepth}{1}}

\renewcommand{\theequation}{\thesection.\arabic{equation}}

\section{Definitions of the Hadronic Matrix Elements}
\label{app:definitions}
\setcounter{equation}{0}

Following Ref.~\cite{Bobeth:2017vxj}, we use a common set of Lorentz structure $\SP{}{\lambda}$
to decompose both the local and the non-local hadronic matrix elements emerging in
$B \to P$(seudoscalar) and $B\to V$(ector) transitions.
In this work, we restrict ourselves to the $B\to K$, $B \to K^*$, and $B_s\to \phi$ transitions, but the considerations of this appendix also apply to, e.g., the $B_s\to K$ and $B_s \to K^*$ transitions.

\subsection{$B\to P$ Transitions}

For $B\to P$ transitions there are two independent Lorentz structures $\SP{\mu}{\lambda}$ in the decomposition
of the matrix elements. Here $\mu$ is a Lorentz index and $\lambda = 0,t$ denotes either longitudinal or
timelike polarization of the underlying current.
The structures read
\begin{align}
\label{eq:def:SP}
    \SP{\mu}{0}(k, q) & \equiv 2 k_\mu - \frac{2 (q\cdot k)}{q^2} q_\mu \,, &
    \SP{\mu}{t}(k, q) & \equiv \frac{M_B^2 - M_P^2}{q^2} q_\mu\,.
\end{align}
Using these structures, we decompose the $B\to P$ matrix elements into form factors
as follows:
\begin{align}
    \label{eq:def:hel-ff-v-BK}
    \FP{\mu}(k, q)
        & \equiv 
    \bra{P(k)} \bar{s}\gamma_\mu P_L\,b \ket{\bar{B}(q+k)} = \frac{1}{2}\left[ \SP{\mu}{0} \, \FP{0} + \SP{\mu}{t} \, \FP{t} \right] \,,
    \\
    \label{eq:def:hel-ff-t-BK}
    \FP{T,\mu}(k, q)
        & \equiv 
    \bra{P(k)} \bar{s}\sigma_{\mu\nu} q^\nu P_R\,b\ket{\bar{B}(q+k)}
        = \frac{i}{2}\, M_B \,  \SP{\mu}{0} \, \FP{0,T}
        \,,\\
    \HP{\mu}(k, q)
        & \equiv i\,\int d^4x\, e^{iq\cdot x}
                 \bra{P(k)} T\big\{ j^\text{em}_\mu(x), (C_1\cO_1 + C_2\cO_2)(0) \big\} \ket{\bar{B}(q+k)} 
                 \nonumber\\*
    \label{eq:def:nonlocBK}
        & = M_B^2\, \SP{\mu}{0}\, \HP{0} \,.
\end{align}
Here and throughout this article we suppress  the argument of the local and non-local form factors, which are functions of the momentum transfer squared: $\FM{\lambda} \equiv \FM{\lambda}(q^2)$ and $\HM{\lambda} \equiv \HM{\lambda}(q^2)$. 

The relations between our local form factor basis and the traditional basis of form factors (see, e.g., Refs.~\cite{Gubernari:2018wyi,Khodjamirian:2006st}) read
\begin{align}
    \label{eq:rel:BK-ff-to-us}
    \FP{0}        & = f_+^{B\to P}, &
    \FP{t}        & = f_0^{B\to P}, & 
    \FP{0,T}      & = \frac{q^2}{M_B(M_B+M_K)} f_T^{B\to P} \,.
\end{align}
The non-local form factor \HP{0} defined in \Eq{eq:def:nonlocBK} is related to the non-local form factor \HP{ } defined in~\Reff{Khodjamirian:2010vf} through
\begin{align}
    \HP{0} = - Q_c \frac{q^2}{2 M_B^2} \HP{ }
    \,.
\end{align}

\subsection{$B\to V$ Transitions}

For $B\to V$ transitions there are four independent Lorentz structures $\SV{\alpha\mu}{\lambda}$ in the decomposition
of the matrix elements. Here $\alpha$ and $\mu$ are Lorentz indices and $\lambda = \perp,\para,0,t$ denotes the different polarization of the underlying current.
The structures read
\begin{equation}
\begin{aligned}
    \label{eq:def:S}
    \SV{\alpha\mu}{\perp}(k, q) & = \frac{\sqrt{2}\, M_B}{\sqrt{\lamkin}} \epsilon_{\alpha\mu k q}\,, &
    \SV{\alpha\mu}{\para}(k, q) & = \frac{i\, M_B}{\sqrt{2}}\left[g_{\alpha\mu} - \frac{4(q\cdot k)}{\lamkin} q_\alpha k_\mu
                          + \frac{4 M_V^2}{\lamkin} q_\alpha q_\mu\right]\,, \\
    \SV{\alpha\mu}{t}(k, q)     & = \frac{2i\, M_V}{q^2} q_\alpha q_\mu\,, &
    \SV{\alpha\mu}{0}(k, q)     & = \frac{4i\, M_V M_B^2}{q^2 \lamkin}
                            \left[q^2 q_\alpha k_\mu - (q\cdot k)\,q_\alpha q_\mu\right]\,.
\end{aligned}    
\end{equation}
where $\lamkin \equiv \lambda(M_B^2,M_V^2,q^2)$ is the K\"all\'en function.
We decompose the local and non-local $B\to V$ matrix elements as
\begin{align}
    \, \FV{\mu}(k, q)
    & \equiv 
     \bra{V(k, \eta)} \bar{s}\gamma_\mu P_L \,b \ket{\bar{B}(q+k)}
    \nonumber\\
        & =\frac{1}{2}\, \eta^{*\alpha} \left[ \SP{\alpha\mu}{\perp} \FV{\perp}
        -  \SP{\alpha\mu}{\para} \FV{\para} - \SP{\alpha\mu}{0} \FV{0} - \SP{\alpha\mu}{t} \FV{t}\right]
    \,,\\
    \hspace{-0.5cm} 
    \, \FV{T,\mu}(k, q)
    & \equiv
     \bra{V(k, \eta)} \bar{s}\sigma_{\mu\nu} q^\nu P_R\, b\ket{\bar{B}(q+k)}
   \nonumber \\ 
    &= \frac{i}{2}\, M_B\, \eta^{*\alpha}\, \left[\SP{\alpha\mu}{\perp} \FV{\perp,T}
    -\SP{\alpha\mu}{\para} \FV{\para,T} - \SP{\alpha\mu}{0} \FV{0,T}\right]\,,\\
    \label{eq:def:correlator-BKstar}
    \HV{\mu}(k, q)
        & \equiv i\,\int d^4x\, e^{iq\cdot x}
                 \bra{V(k, \eta)} T\big\{ j^\text{em}_\mu(x), (C_1\cO_1 + C_2\cO_2)(0) \big\} \ket{\bar{B}(q+k)}
        \nonumber\\
        & = M_B^2\, \eta^{*\alpha}\,\left[
        \SP{\alpha\mu}{\perp} \HV{\perp} - \SP{\alpha\mu}{\para} \HV{\para} - 
        \SP{\alpha\mu}{0} \HV{0}\right]\,.
\end{align}
The minus signs in front of \FV{\para} and \FV{0} have been introduced
to ensure that the form factors are all positive in the semileptonic phase space;
the signs in front of $\HV{\para}$ and $\HV{0}$ then follow.
The relations between our local form factor basis and the traditional basis of form factors (see, e.g., Refs.~\cite{Gubernari:2018wyi,Khodjamirian:2006st}) read
\begin{equation}
\begin{aligned}
    \FV{\perp}        & = \frac{\sqrt{2\,\lamkin}}{M_B (M_B + M_V)} V\,, &
    \FV{\para}        & = \frac{\sqrt{2}\,(M_B + M_V)}{M_B} A_1\,,       \\
    \FV{0}            & = \frac{(M_B^2 - M_V^2 - q^2)(M_B + M_V)^2 A_1 - \lamkin A_2}{2 M_V M_B^2 (M_B + M_V) }\,, &
    \FV{t}            & = A_0\,,\\
    \FV{\perp,T}  & = \frac{\sqrt{2\, \lamkin}}{M_B^2} T_1\,, &
    \FV{\para,T}  & = \frac{\sqrt{2} (M_B^2 - M_V^2)}{M_B^2} T_2\,, 
\end{aligned}   
\end{equation}
\vspace{-0.5cm}
\begin{align*}
    \!\!\!\!\!\!\!\!\!\!\!\!\!\!\!\!\!\!\!\!\!\!\!\!\!\!\!\!\!\!\!\!\!\!\!\!\!\!\!\!\!\!\!\!\!\!\!
    \FV{0,T}      & = \frac{q^2 (M_B^2 + 3 M_V^2 - q^2)}{2 M_B^3 M_V} T_2
                                - \frac{q^2\lamkin}{2 M_B^3 M_V (M_B^2 - M_V^2)} T_3\,.
\end{align*}
The non-local form factor $\HV{0}$ defined in \Eq{eq:def:correlator-BKstar} is related to the non-local form factor \HV[]{i} defined in Ref.~\cite{Khodjamirian:2010vf} through
\begin{equation}
    \label{eq:rel:H_i-to-us}
    \begin{aligned}
    \HV{\perp}  & = Q_c\, \frac{\sqrt{\lamkin}}{\sqrt{2} M_B^3} \, \HV[]{1}\,,       \\
    \HV{\para} & = -\sqrt{2}\,Q_c\,\frac{M_B^2 - M_V^2}{M_B^3} \,  \HV[]{2}\,, \\
    \HV{0}     & =
    -Q_c\,\frac{q^2}{2 M_B^4 M_V}\, 
                                \left[(M_B^2 + 3 M_V^2 - q^2)\HV[]{2} - \frac{\lamkin}{M_B^2 - M_V^2} \HV[]{3} \right]\,.
    \end{aligned}
\end{equation}

\section{Matching Coefficient at Subleading Power}
\label{app:Itensor}
\setcounter{equation}{0}

The matching coefficient for the next-to-leading power of the LCOPE of correlator (\ref{eq:LOLCOPE}) was computed for the first time in Ref.~\cite{Khodjamirian:2010vf}.
The result was written in the form
\begin{multline}
    \label{eq:Itensor}
    I_{\mu\rho\alpha\beta}(q,\omega_2)=\frac{1}{8 \pi^2}\int_0^1 du \Bigg\{
    \left[
    \bar{u}\tilde{q}_\mu\tilde{q}_\alpha g_{\rho\beta}+
    u      \tilde{q}_\rho\tilde{q}_\alpha g_{\mu\beta}-
    \bar{u}\tilde{q}^2 g_{\mu\alpha} g_{\rho\beta}
    \right] \frac{dI(\tilde{q}^2)}{d\tilde{q}^2}
    \\
    -\frac{\bar{u}-u}{2} g_{\mu\alpha} g_{\rho\beta}I(\tilde{q}^2)
    \Bigg\}
    \,,
\end{multline}
where
\begin{align*}
    I(\tilde{q}^2)&=\int_0^1 dt \ln\left[
    \frac{\mu^2}{m_c^2-t(1-t)\tilde{q}^2}
    \right],\\
    \tilde{q}^\mu& = q^\mu - v^\mu u\omega_2
    \,.
\end{align*}
It is convenient to rewrite the function $I(\tilde{q}^2)$ as
\begin{equation}
\begin{aligned}
    I(\tilde{q}^2)
        & = \int_0^1 dt\, \ln\left[\frac{\mu^2}{m_c^2 - t(1-t) \tilde{q}^2}\right]\\
        & = \left\lbrace t\cdot \ln \left[\frac{\mu^2}{m_c^2 - t(1-t) \tilde{q}^2}\right]  \right\rbrace_{t=0}^{t=1}
          - \int_0^1 dt\, \frac{\tilde{q}^2\, t(1-2t)}{m_c^2 - t(1-t) \tilde{q}^2}\\
        & = \ln\left[\frac{\mu^2}{m_c^2}\right] - \int_0^1 dt\, \frac{\tilde{q}^2\, t(1-2t)}{m_c^2 - t(1-t) \tilde{q}^2}\,.
\end{aligned}
\end{equation}
Since the part of $I_{\mu\rho\alpha\beta}$ proportional to $I(\tilde{q}^2)$ is multiplied by $(\bar{u}-u)$, the term $\ln\left[\frac{\mu^2}{m_c^2}\right] $ vanishes after integrating over $u$. In addition, using the identity
\begin{align}
    \int_0^1 dt\, \frac{t}{m_c^2 - t(1-t) \tilde{q}^2}=    \int_0^1 dt\, \frac{1}{2}\frac{1}{m_c^2 - t(1-t) \tilde{q}^2} \,,
\end{align}
one obtains
\begin{align}
    \label{eq:Iq2}
    I(\tilde{q}^2)
        & = \int_0^1 dt\,\left[ -\frac{\tilde{q}^2\, t(1-t)}{m_c^2 - t(1-t) \tilde{q}^2}  
        +\frac{\tilde{q}^2}{2}\frac{1}{m_c^2 - t(1-t) \tilde{q}^2} \right]\,.
\end{align}
The calculation of the first derivative of $I(\tilde{q}^2)$ is straightforward:
\begin{align}
    \label{eq:dIq2}
    \frac{dI(\tilde{q}^2)}{d\tilde{q}^2}
        & = \int_0^1 dt\, \frac{
        t(1-t)}{m_c^2 - t(1-t) \tilde{q}^2}\,.
\end{align}
Exploiting \Eqs{eq:Iq2}{eq:dIq2} and absorbing the Levi-Civita symbol coming from $\tilde{G}^{\alpha\beta} \equiv \frac{1}{2} \eps^{\alpha\beta}{}_{\sigma\tau} G^{\sigma\tau}$, we obtain
\begin{align}
    \tilde{I}_{\mu\rho\sigma\tau} (q, \omega_2) 
    &\equiv
    \frac{1}{2}
    \eps^{\alpha\beta}{}_{\sigma\tau}
    I_{\mu\rho\alpha\beta}(q, \omega_2)
        =  \int_0^1 du \int_0^1 dt\,
        \frac{1}{64 \pi^2 ( t(1-t) \tilde{q}^2 - m_c^2)}
        \nonumber\\
    &\times \bigg(
                4 t(1-t) \left( \tilde{q}_\mu \eps_{\rho \sigma\tau \lbrace \tilde{q}\rbrace} 
                - 2 u \tilde{q}_\tau \eps_{ \mu \rho \sigma \lbrace \tilde{q}\rbrace}
                + 2 u \tilde{q}^2 \eps_{\mu \rho \sigma\tau}
                \right)
                + \tilde{q}^2 \left(1 - 2u\right) \eps_{\mu \rho \sigma\tau}
            \bigg)\,.
\end{align}
In this work we adopt the convention $\eps_{0123} =+1$.

\section{Outer Functions}
\label{app:fs}
\setcounter{equation}{0}

To cast the dispersive bound into the form of \Eq{eq:disp-rel-z}, we define the outer functions such that their moduli squared coincide with the weight factors in \Eq{eq:dispbound} on the  integration domain:
\begin{align}
\label{eq:weightBK}
    \left|\outerF[B\to P]{0}(z(\alpha))\right|^2
    & \equiv
    \frac{1}{\chi^\OPE(Q^2)}\frac{8 \pi^2}{3}
    \left|\frac{dz(\alpha)}{d\alpha} \frac{ds(z)}{dz}\right| \frac{M_B^4\,\lambda^{3/2}(M_B^2, M_P^2, s)}{s^4 (s-Q^2)^3}\, \Bigg|_{\scriptsize \begin{array}{c}s=s(z)\\[-1mm] z=z(\alpha)\end{array}} 
    \ ,
    \\
    \left|\outerF[B\to V]{\perp,\para}(z(\alpha))\right|^2
    & \equiv
    \frac{1}{\chi^\OPE(Q^2)}\frac{8 \pi^2}{3}
    \left|\frac{dz(\alpha)}{d\alpha} \frac{ds(z)}{dz}\right| \frac{M_B^6\,\sqrt{\lambda(M_B^2, M_V^2, s)}}{s^3 (s-Q^2)^3} \Bigg|_{\scriptsize \begin{array}{c}s=s(z)\\[-1mm] z=z(\alpha)\end{array}}
    \,,
    \\
    \label{eq:weightBsphi}
    \left|\outerF[B\to V]{0}(z(\alpha))\right|^2
    & \equiv
    \frac{1}{\chi^\OPE(Q^2)}\frac{8 \pi^2}{3}
    \left|\frac{dz(\alpha)}{d\alpha} \frac{ds(z)}{dz}\right| \frac{M_B^8\,\sqrt{\lambda(M_B^2, M_V^2, s)}}{s^4 (s-Q^2)^3} \Bigg|_{\scriptsize \begin{array}{c}s=s(z)\\[-1mm] z=z(\alpha)\end{array}}
    \,.
\end{align}
Since these weight factors contain singularities,
the outer functions should also be defined such that they do not exhibit kinematical singularities on the open unit disk of the $z$ plane~\cite{Caprini:2019osi}.
We only retain such kinematical singularities at $z(s = 0)$ that ensure the correct physical behaviour of the non-local form factors $\HM{\lambda}$ involving an on-shell photon, i.e.~the absence of an on-shell longitudinal photon.
Specifically, we keep a $1/s(z)$ pole in both $\outerF[B\to P]{0}$ and $\outerF[B\to V]{0}$.
In this way, our parametrization for the non-local form factors $\HM{\lambda}$, i.e.
\begin{align}
    \HM{\lambda} = \frac{1}{\outerF[B\to M]{\lambda}(z) \, \P(z)}
    \sum_{n} a_{\lambda,n}^{B\to M} p_{n}^{B\to M}(z)
    \,,
\end{align}
explicitly satisfies the conditions $\HP{0}(s=0)=0$ and $\HV{0}(s=0)=0$.
\\

\begin{table}[t!]
    \newcommand{\pp}{\phantom{+}}
    \centering
    \renewcommand{\arraystretch}{0.90}
    \begin{tabular}{ccccc}
        \toprule
        \multirow{2}{*}{Outer function}               &
        \multicolumn{4}{c}{Parameters}
                                 \\
                                 & 
        $a$                      &
        $b$                      &
        $c$                      &
        $d$                      \\
        \toprule
        \rule{0pt}{17pt}
        $\outerF[B\to P]{0}$     & 
        3                        &
        3                        &
        2                        &
        2                        \\
        \rule{0pt}{17pt}
        $\outerF[B\to V]{\perp}=
        \outerF[B\to V]{\para}$  & 
        3                        &
        1                        &
        3                        &
        0                        \\
        \rule{0pt}{17pt}
        $\outerF[B\to V]{0}$     & 
        3                        &
        1                        &
        2                        &
        2                        \\[7pt]
        \bottomrule
    \end{tabular}
    \caption{ \it
    \label{tab:outer} 
      Parameters of the outer functions \Eq{eq:outermaster}.
    }
\end{table}

The outer function are then written as
\eqa{
\label{eq:outermaster}
\outerF[B\to M]{\lambda}(z) & = & \N_\lambda
(1+z)^{\frac{1}{2}}
(1-z)^{a-b+c+d-\frac{3}{2}}
\phi_1(z)^a \phi_2(z)^{\frac{b}{2}} \phi_3(z)^c \phi_4(z)^d \ ,
}
where we have defined the constant factor
\begin{align}
    \label{eq:outerconst}
    \N_\lambda & = 
    4 \pi M_B^{a-b+c+d-2}
    \sqrt{\frac{2(4M_D^2-s_0)}{3\,\chi^\OPE(Q^2)}}\,,
\end{align}
and the functions 
\begin{align}
    \phi_1(z) & =
    -\frac{\left(2 \sqrt{\left(4 M_D^2-Q^2\right) \left(4
   M_D^2-s_0\right)}+8 M_D^2-Q^2-s_0\right)^{\frac{1}{2}}}{2 \sqrt{\left(4
   M_D^2-Q^2\right) \left(4 M_D^2-s_0\right)}+8
   M_D^2+Q^2 (z-1)-s_0 (z+1)}
    \,,\\
    \phi_2(z) & =
    \Big(M_B^4 (z-1)^4-2 M_B^2 (z-1)^2 \left(-16 M_D^2
   z+M_M^2 (z-1)^2+s_0 (z+1)^2\right)
   \nonumber\\ &
   +\left(16
   M_D^2 z+M_M^2 (z-1)^2-s_0 (z+1)^2\right)^2\Big)^{\frac{1}{2}}
    \,,\\
    \phi_3(z) & =
    \frac{\left(8 M_D^2+4 \sqrt{4 M_D^4-M_D^2 s_0}-s_0\right)^{\frac{1}{2}}}{-8
   M_D^2-4 \sqrt{4 M_D^4-M_D^2 s_0}+s_0(z+1)}
    \,,\\
    \phi_4(z) & =
    \left(s_0 (z+1)^2-16 M_D^2 z\right)^{-\frac{1}{2}}
    \,.
\end{align}
Here $s_0$ is a parameter of the $s\to z$ mapping (\ref{eq:zdef}).
The values of the parameters $a,\,b,\,c,\,d$ of \Eq{eq:outermaster} for the outer functions considered in this work are listed in \Tab{tab:outer}.

\section{Local $B_s \to \phi$ Form Factors}
\label{app:BsPhiFFs}

We compute the local $B_s \to \phi$ form factors using the analytical results of Ref.~\cite{Gubernari:2018wyi}.
We provide our numerical results at $q^2=\{-15,-10,-5,0,+5\}\,\GeV^2$ through a machine-readable YAML file attached to the arXiv preprint of this article as an ancillary file.
We use the same format as in Ref.~\cite{Bordone:2019guc}.
The inputs used to obtain these results are discussed in \SubSec{sec:numres}.

In \Tab{tab:FFscomp}, we compare our results at $q^2=0$, obtained by means of LCSRs with $B$-meson LCDAs, with the results of Ref.~\cite{Straub:2015ica}, obtained by means of LCSRs with light-meson LCDAs.
We find perfect agreement between these two calculations. The larger uncertainties in our calculation with respect to Ref.~\cite{Straub:2015ica} are due to the fact that the $B$-meson LCDAs are not as well known as the light-meson LCDAs.

\begin{table}[t!]
    \newcommand{\pp}{\phantom{+}}
    \centering
    \renewcommand{\arraystretch}{1.45}
    \begin{tabular}{cccc}
        \toprule
        Local $B_s \to \phi$ form factors  & 
        This work                          & 
        Ref.~\cite{Straub:2015ica}         \\
        \toprule
        $V(q^2 = 0)$                       & 
        $0.387\pm0.111$                       &
        $0.387\pm0.033$                      \\
        $A_0(q^2 = 0)$                       & 
        $0.372\pm0.070$                       &
        $0.389\pm0.045$                     \\
        $A_1(q^2 = 0)$                       & 
        $0.304\pm0.080$                       &
        $0.296\pm0.027$                       \\
        $T_1(q^2 = 0)$                       & 
        $0.339\pm0.093$                       &
        $0.309\pm0.027 $                      \\
        $T_{23}(q^2 = 0)$                       & 
        $0.651\pm0.115$                       &
        $0.676\pm0.071$                       \\
        \bottomrule
    \end{tabular}
    \caption{ \it
    \label{tab:FFscomp} 
      Comparison between the local form factors results of Ref.~\cite{Straub:2015ica} and our results at $q^2 = 0$.
    }
\end{table}

\newpage
 
\bibliographystyle{JHEP}
\bibliography{references}

\end{document}